\documentclass[sigconf]{acmart}
\copyrightyear{2026}
\acmYear{2026}
\setcopyright{cc}
\setcctype{by-nc-nd}
\acmConference[DIS '26]{Designing Interactive Systems Conference}{June 13--17, 2026}{Singapore, Singapore}
\acmBooktitle{Designing Interactive Systems Conference (DIS '26), June 13--17, 2026, Singapore, Singapore}
\acmDOI{10.1145/3800645.3812990}
\acmISBN{979-8-4007-2563-0/2026/06}

\usepackage[shortlabels]{enumitem}

\newenvironment{quot}[2]{
\begin{quote}
\begin{center}
\hspace{1em}~\textit{``#1''}\hfill\\
~\textit{---}~\faCommentingO~\textit{#2}~\hspace{1em}
\end{center}
\end{quote}
}

\usepackage{xspace}

\newcommand{\varname}[1]{\texttt{#1}}
\newcommand{\sig}[1]{\cellcolor{palepink!50}\textcolor{red}{#1}}
\newcommand{\pattern}[1]{\textit{#1 pattern}}

\usepackage{multirow}
\usepackage{fontawesome}

\makeatletter
\newdimen\cdl@dashwidth  \cdl@dashwidth=4pt   
\newdimen\cdl@dashgap    \cdl@dashgap=4pt     
\newcount\cdl@cnta                            
\newcount\cdl@cntb                            

\def\cdl@hline{%
  \vrule \@height \arrayrulewidth \@width \cdl@dashwidth
  \hskip.5\cdl@dashgap
  \xleaders\hbox{\hskip.5\cdl@dashgap
    \vrule \@height \arrayrulewidth \@width \cdl@dashwidth
    \hskip.5\cdl@dashgap}\hfill
  \hskip.5\cdl@dashgap
  \vrule \@height \arrayrulewidth \@width \cdl@dashwidth
  \cr
  \noalign{\vskip-\arrayrulewidth}%
}

\newcommand{\cdashlinelr}[1]{\cdl@parse#1\@nil}
\def\cdl@parse#1-#2\@nil{%
  \noalign{%
    \vskip\aboverulesep
    \global\cdl@cnta#1\relax
    \global\advance\cdl@cnta\m@ne
    \ifnum\cdl@cnta>\z@
      \global\let\@gtempa\cdl@a
    \else
      \global\let\@gtempa\cdl@b
    \fi
    \global\cdl@cntb#2\relax
    \global\advance\cdl@cntb-\cdl@cnta
  }%
  \@gtempa
  \noalign{\vskip\belowrulesep}%
}
\def\cdl@a{\multispan\cdl@cnta&\multispan\cdl@cntb\unskip\cdl@hline}
\def\cdl@b{\multispan\cdl@cntb\unskip\cdl@hline}
\makeatother

\usepackage{colortbl}
\definecolor{palered}{HTML}{b8778d}
\definecolor{paleorange}{HTML}{ffd6a5}
\definecolor{paleyellow}{HTML}{fdffb6}
\definecolor{palegreen}{HTML}{93d275}
\definecolor{palecyan}{HTML}{9bf6ff}
\definecolor{paleblue}{HTML}{a2d2ff}
\definecolor{palepurple}{HTML}{CDB4DB}
\definecolor{actualpurple}{HTML}{560BAD}
\definecolor{palepink}{HTML}{f7c9d8}
\definecolor{palewhite}{HTML}{fffffc}
\definecolor{palegray}{HTML}{e5e5e2}
\definecolor{brickred}{HTML}{C9030A}

\definecolor{revisioncolor}{HTML}{000000}  

\newcommand{\revision}[1]{\textcolor{revisioncolor}{#1}}  

\definecolor{tagFH}{HTML}{1D4ED8}   
\definecolor{tagFHbg}{HTML}{DBEAFE}
\definecolor{tagBD}{HTML}{9333EA}   
\definecolor{tagBDbg}{HTML}{F3E8FF}
\definecolor{tagPZ}{HTML}{DC2626}   
\definecolor{tagPZbg}{HTML}{FEE2E2}
\definecolor{tagHI}{HTML}{EA580C}   
\definecolor{tagHIbg}{HTML}{FFEDD5}
\definecolor{tagFR}{HTML}{0D9488}   
\definecolor{tagFRbg}{HTML}{CCFBF1}
\definecolor{tagPNF}{HTML}{059669}  
\definecolor{tagPNFbg}{HTML}{D1FAE5}
\definecolor{tagET}{HTML}{B45309}   
\definecolor{tagETbg}{HTML}{FEF3C7}
\definecolor{tagPM}{HTML}{7C3AED}   
\definecolor{tagPMbg}{HTML}{EDE9FE}
\definecolor{tagCL}{HTML}{0369A1}   
\definecolor{tagCLbg}{HTML}{E0F2FE}
\definecolor{tagTQ}{HTML}{BE185D}   
\definecolor{tagTQbg}{HTML}{FCE7F3}
\definecolor{tagNG}{HTML}{CA8A04}   
\definecolor{tagNGbg}{HTML}{FEF9C3}
\definecolor{tagNone}{HTML}{484d56} 
\definecolor{tagNonebg}{HTML}{F3F4F6}

\newcommand{\ptag}[3]{%
  \colorbox{#1}{\textcolor{#2}{\sffamily\small%
    \rule[-0.20em]{0pt}{0.75em}#3}}%
}

\newcommand{\tagFH}{\ptag{tagFHbg}{tagFH}{False Hierarchy/Visual Prominence}}
\newcommand{\tagBD}{\ptag{tagBDbg}{tagBD}{Bad Default}}
\newcommand{\tagPZ}{\ptag{tagPZbg}{tagPZ}{Privacy Zuckering}}
\newcommand{\tagHI}{\ptag{tagHIbg}{tagHI}{Hidden Information}}
\newcommand{\tagFR}{\ptag{tagFRbg}{tagFR}{Forced Registration}}
\newcommand{\tagPNF}{\ptag{tagPNFbg}{tagPNF}{Positive or Negative Framing}}
\newcommand{\tagET}{\ptag{tagETbg}{tagET}{Endorsements and Testimonials}}
\newcommand{\tagPM}{\ptag{tagPMbg}{tagPM}{Privacy Maze}}
\newcommand{\tagCL}{\ptag{tagCLbg}{tagCL}{Complex \& Lengthy Language}}
\newcommand{\tagTQ}{\ptag{tagTQbg}{tagTQ}{Trick Questions}}
\newcommand{\tagNG}{\ptag{tagNGbg}{tagNG}{Nagging}}
\newcommand{\tagNone}{\ptag{tagNonebg}{tagNone}{None}}

\begin{document}

\title[User Experiences and Responses to VR Privacy Deceptive Design]{Rushed by Discomfort, Trapped by Immersion: Users' Experiences and Responses to Privacy Deceptive Design in Commercial VR Applications}


\author{Hilda Hadan}
\email{hhadan@uwaterloo.ca}
\orcid{https://orcid.org/0000-0002-5911-1405}
\affiliation{
    \institution{Stratford School of Interaction Design and Business, University of Waterloo}
    \city{Waterloo}
    \country{Canada}
}

\author{Michaela Valiquette}
\email{mjvaliqu@uwaterloo.ca}
\orcid{https://orcid.org/0009-0009-5335-3014}
\affiliation{
    \institution{Stratford School of Interaction Design and Business, University of Waterloo}
    \city{Waterloo}
    \country{Canada}
}

\author{Lennart E. Nacke}
\email{lennart.nacke@acm.org}
\orcid{https://orcid.org/0000-0003-4290-8829}
\affiliation{
    \institution{Stratford School of Interaction Design and Business, University of Waterloo}
    \city{Waterloo}
    \country{Canada}
}

\author{Leah Zhang-Kennedy}
\email{lzhangke@uwaterloo.ca}
\orcid{https://orcid.org/0000-0002-0756-0022}
\affiliation{
    \institution{Stratford School of Interaction Design and Business, University of Waterloo}
    \city{Waterloo}
    \country{Canada}
}

\renewcommand{\shortauthors}{Hadan et al.}

\begin{abstract}
Commercial Virtual Reality (VR) transforms people's virtual experiences but introduces deceptive design opportunities that threaten user privacy. 
Although privacy deceptive patterns on 2D platforms are well-documented, their impacts in VR remain understudied.
We surveyed 481 users' experiences and responses to privacy deceptive patterns across eight commercial VR scenarios.
We found that VR deceptive design can exploit both cognitive vulnerabilities and bodily strain, \revision{a phenomenon we define as \textit{Ergonomic Susceptibility},} and that VR's sensory-rich experiences can make users more likely to accept invasive data disclosure framed as immersion-preserving. Users recognized manipulation but their prior non-VR exposure can foster privacy resignation.
Our study shows ergonomics is a critical factor in future privacy-preserving VR design, and urges VR researchers, designers, and policymakers to develop ethical design and privacy management solutions that account for VR's unique multimodal, immersive, and ergonomic properties, building immersive experiences that respect user privacy and mitigate manipulative data practices. 




\end{abstract}




\maketitle

\section{Introduction}
\label{sec:introduction}


Virtual Reality (VR) technologies are transforming how people work, play, and socialize through unbounded display, realistic simulation, and multi-sensory feedback~\cite{slater2009place,skarbez2021revisiting}. However, VR properties that enable immersive experiences also introduce novel opportunities for deceptive design\footnote{We follow the ACM Diversity and Inclusion Council's guideline for inclusive language and adopt the term ``deceptive design'' instead of ``dark patterns'' in our study. See: \url{https://www.acm.org/diversity-inclusion/words-matter}} and intensify \revision{user privacy risks.}  
\revision{Privacy} deceptive design refers to practices that undermine users' ability to make voluntary and informed decisions about their information, regardless of the designer's intent~\cite{gray2024ontology,bosch2016tales}. \revision{In VR, polished simulations and disconnection from physical surroundings can distort user perception~\cite{tseng2022perceptual,mhaidli2021identifying,buck2022security}, lead to involuntary interactions~\cite{tseng2022perceptual,liu2025flytrap,su2022perception}, and enable detailed behavioral profiling~\cite{mhaidli2021identifying,Greenberg2014proxemic}, amplifying the potential for manipulation~\cite{bonnail2022exploring,hadan2024deceived,krauss2024makes}.} Despite these risks, little is known about how VR end-users experience, interpret, and respond to privacy deceptive design in commercial VR applications. Without this understanding, effective risk communication cannot be created, leaving users vulnerable to persuasive data exploitation and undermining the public trust essential for VR's mainstream adoption~\cite{garg2012end,abraham2022implications,hadan2024privacy}. 

Deceptive design has become an important focus in privacy research for its ability in steering users away from informed decisions and toward actions that compromise their privacy~\cite{Brignull2023book,bosch2016tales}. Common privacy deceptive tactics include pre-enabled data-sharing settings, easy-to-register but difficult-to-delete user accounts, and bundled consent that pressure users to accept all options~\cite{gunawan2022redress,luguri2021shining}. Existing privacy research on deceptive design has primarily focused on identifying such patterns in 2D computer and mobile environments (e.g.,~\cite{gunawan2021comparative,soe2020circumvention,nouwens2020dark}). Studies of privacy deceptive design in VR contexts have largely relied on expert predictions, controlled lab studies, and literature-based speculation (e.g.,~\cite{hadan2024deceived,krauss2024makes,bonnail2022exploring}). While previous work provides a classification of deceptive patterns in commercial VR applications from an expert perspective~\cite{hadan2024cscw}, how VR end-users actually are influenced by these patterns remains an open and critical question. Our paper addresses this gap by providing the first empirical study of users' experiences and responses to privacy deceptive patterns in commercial VR applications.

Our research aims to understand \textit{how VR end-users experience, interpret, and respond to privacy deceptive designs in commercial VR applications.} Specifically, we study how deceptive design affects users' understanding of data practices, their data control decisions, and how VR's unique properties shape users' responses. Our research questions (RQs) are: 

\begin{enumerate}[label= \textbf{RQ\arabic*:}]
    \item To what extent are users aware of and concerned about privacy impacts of deceptive design in VR?
    \item What factors relate to VR users' perceived benefits and privacy impact when they encounter deceptive design? 
    \item How do VR properties shape users' interpretation and responses to privacy deceptive design?
\end{enumerate}

Building on prior research on user privacy perceptions in immersive technologies~\cite{hadan2024privacy,harborth2021investigating} and deceptive design in computer and mobile environments~\cite{bongard2021definitely,geronimo2020ui,habib2022okay,gray2021end}, we conducted an online survey with $N=481$ VR users. Participants were assigned to one of eight pre-designed scenarios, each illustrating a VR design mechanism derived from a systematic observation of four widely used commercial VR applications they had used previously and were still actively using at the time of our study (see~\autoref{subsec:demographics}). We asked participants to reflect on how they would respond to the VR mechanisms, including their decisions, reactions, and reasoning, drawing on their prior experience with the applications.
This approach allowed us to capture VR users' reflections on privacy deceptive patterns within familiar VR contexts and their interpretation of impact on their privacy understanding and decision-making, while minimizing the ethical risks from directly exposing participants to privacy-invasive and manipulative VR contexts.

Our research extends the Human-Computer Interaction (HCI) literature on privacy deceptive design in VR through three main contributions. \textit{First}, user-perspective of VR hardware-induced discomfort suggests that it can increase users' susceptibility to deceptive patterns involving cumbersome interactions. Our research identifies ergonomics as a critical factor for future VR privacy design, guiding researchers to address how physical discomfort impairs users' privacy decision-making and resistance to deceptive practices. \textit{Second}, we show that the sensory richness of VR can amplify users' vulnerability to deceptive practices that framed data disclosure as sustaining immersive experiences. Our findings highlight the importance of VR design approaches that preserve immersion while preventing its role in normalizing harmful data practices. \textit{Third}, we show that users' prior exposure to deceptive patterns in non-VR contexts and trust in publishers can cause them to downplay privacy risks in VR. Therefore, VR researchers, industry practitioners, and policymakers must extend ethical design approaches and regulatory frameworks to immersive technologies, and create multimodal privacy management solutions that address the unique vulnerabilities arising from VR's immersive properties, multimodal interactions, and ergonomic limitations.

\section{Related Work}
\label{sec:related-work}

We situate our study within existing literature on deceptive design, its VR manifestation, and associated privacy issues. 

\subsection{Deceptive Design in VR and Multi-Sensory Virtual Environments}

Deceptive design has been studied across websites~\cite{Brignull2023book,opc2024sweep,gray2018dark}, mobile applications~\cite{geronimo2020ui,gunawan2021comparative}, games~\cite{hadan2024ow2,king2023investigating}, and 3D immersive virtual environments~\cite{Greenberg2014proxemic,hadan2024deceived,krauss2024makes}. In immersive virtual contexts, recent research has synthesized deceptive practices from the literature~\cite{hadan2024deceived}, and examined expert predictions about future use cases~\cite{krauss2024makes}. In commercial VR, studies found that privacy deceptive designs use traditional interface elements and VR properties to influence users' information disclosure behavior and privacy awareness~\cite{hadan2024computer}, and that manipulative VR advertising can compromise user experience and perception~\cite{mhaidli2025intriguing}. Other studies showed that haptic feedback~\cite{tang2025dark} and electrical muscle stimulation~\cite{liu2025flytrap} can enable novel manipulation forms. AI's potential to introduce new deceptive opportunities has also drawn attention~\cite{krauss2025create}. 


\subsubsection{Unique risks of deceptive design in VR}

Research synthesized two ways VR introduces manipulative tactics~\cite{hadan2024deceived,krauss2024makes}: its immersion, realism, and interactivity; and its capacity to collect granular bodily sensor data. VR's visual, audio, and haptic feedback create asymmetries in users' awareness of their physical surroundings~\cite{speicher2019mixed,skarbez2021revisiting}. Its unbounded displays and real-time interactivity create a strong sense of presence, making users feel physically located in a virtual environment~\cite{slater2009place,skarbez2021revisiting}. This immersive environment allows users to experience otherwise impossible scenarios, such as simulated encounters with deceased relatives~\cite{hayden2020mother}, body swapping~\cite{maister2015changing}, or fear-triggering situations~\cite{bentz2021effectiveness}. Highly polished simulations can lead users to form unrealistic expectations of actual products or experiences~\cite{mhaidli2021identifying}. Tactics such as \revision{perception hacking (e.g., the use of manipulating sensory feedback to alter users' perception of reality and nudge their physical hand and body movements~\cite{tseng2022perceptual})} and modified electrical muscle stimulation can also trigger involuntary interaction with virtual objects~\cite{su2022perception,tseng2022perceptual,liu2025flytrap}. Meanwhile, VR systems collect extensive bodily sensor data, such as gait, posture, and movement patterns, enabling user profiling and even deanonymization~\cite{miller2020personal, pfeuffer2019behavioural}, which can support hyper-personalized simulations based on inferred vulnerabilities or preferences~\cite{buck2022security,bonnail2022exploring}.



\subsection{The Privacy Implications of Deceptive Design}


Deceptive design can compromise users' privacy awareness and decision-making, and cause unintended data disclosures~\cite{zhang2024navigating,bosch2016tales,gunawan2022redress}. Literature has identified deceptive tactics in website and mobile privacy mechanisms~\cite{bosch2016tales,gunawan2022redress,gunawan2021comparative} that create an illusion of informed consent while pressuring users to share personal information for business interests~\cite{waldman2020cognitive}.

\subsubsection{Privacy policy, consent, and notice mechanisms} 

Deceptive design practices frequently appear in privacy mechanisms users encounter when accessing digital services, including consent requests, privacy policies, and data collection notices. Observational research documented widespread deceptive tactics in consent interfaces on major websites~\cite{utz2019uninformed,borberg2022so,sanchezrola2019can}. These tactics include implied or bundled consent, pre-checked boxes, forced actions, visual manipulation, obscured denial options, and omitted terms of service~\cite{gray2021end,gunawan2021comparative,soe2020circumvention}, many of which do not compliant with consumer protection regulations~\cite{gunawan2022redress,trevisan20194,gray2021legal}. User studies show that such tactics increase the likelihood that users consent to data collection they would otherwise reject~\cite{borberg2022so,nouwens2020dark}, and that users often regret these decisions once they understand their implications~\cite{machuletz2019multiple}. 
In privacy policy contexts, deceptive design often appears as obscure hyperlink placement, overly positive framing, and complex legal language that hinder user comprehension~\cite{jensen2004privacy,linden2018privacy,adjerid2013sleights}. In certain cases, users have to provide consent before having access to the full policy, or the policies omit critical data practices, making it difficult for users to understand how their information will be used at the time of consent~\cite{adjerid2013sleights,linden2018privacy,zimmeck2019maps}.

\subsubsection{Registration, subscription, settings, and opt-out mechanisms} 

In registration and subscription processes, deceptive design can manifest as bad default, forced enrollment and continuation, and social engineering to trick users into providing personal and payment information upfront and to create obstacles that keep them subscribed to services they do not use~\cite{sheil2024staying,bosch2016tales,mathur2019dark}. Beyond initial consent, studies found privacy settings that impose additional labors on users through hidden or ineffective controls and bad defaults that discourage modification and weaken privacy protections after initial consent~\cite{geronimo2020ui,gunawan2021comparative,mildner2021ethical}. Research on post-registration also documented missing or delayed cancellation options and user-flow friction that impedes account deletion or consent withdrawal, leaving users involuntarily subscribed or engaged~\cite{geronimo2020ui,gray2025getting,schaffner2022understanding}. Studies of opt-out mechanisms, such as email subscriptions, targeted advertising, data deletion, and Do Not Sell My Private Information (DNSMPI) requests, indicate that these controls are often hard to find, poorly labeled, or ineffective~\cite{California2020CPRA,gunawan2021comparative,van2022setting,tran2025dark}, limiting users' ability to exercise autonomy over their personal data~\cite{habib2020scavenger,habib2019empirical}.

\subsubsection{General user interaction and design mechanisms}
\label{subsubsec:related-work-privacyuserinteraction}

Beyond dedicated privacy mechanisms, VR simulations rely on a wide range of bodily and environmental sensors, such as optical cameras, depth sensors, eye and motion trackers, that produce highly granular data about users and their surroundings~\cite{adams2018ethics,egliston2021examining}. While some of these sensors exist in non-immersive technologies~\cite{hadan2024privacy,o2016convergence}, VR sensors enable a more granular capture of users' full-body representations and interactions with their environment~\cite{adams2018ethics,egliston2021examining,hadan2024privacy}, which can support more accurate inferences about users' behavioral patterns, habits, and cognitive and emotional responses~\cite{pfeuffer2019behavioural,miller2020personal,mhaidli2021identifying}. Deceptive design in VR can extend beyond interface-level design to include tactics such as memory manipulation, blind-spot tracking, and emotional exploitation through realistic simulations~\cite{bonnail2022exploring,mhaidli2021identifying,miller2020personal}. The immersive, social experiences of VR further intensify these risks, especially for vulnerable populations such as children, individuals with disabilities, and users with limited digital literacy~\cite{rossi2024vulnerable,maloney2020anonymity}.


\revision{In summary, prior studies have theorized the unique risks of VR's immersive environments and granular sensor data~ \cite{krauss2024makes,su2022perception,bonnail2022exploring,pfeuffer2019behavioural}. Our research extends these expert- and theory-driven studies with an empirical examination on how VR end-users interpret, respond to, and reflect on privacy deceptive designs in commercial VR applications.}

\section{Methodology}
\label{sec:methodology}
Upon Research Ethics Board's approval, we surveyed $N=481$ VR users on Prolific\footnote{Prolific.~\url{https://www.prolific.com/}}. Eligible participants were at least 18 years old and had prior experience with the commercial VR applications in their assigned scenario. To prevent bias from repeated exposure, we assigned each participant to only one of eight pre-designed scenarios. We analyzed the data using statistical tests and thematic analysis~\cite{clarke2021thematic}. 
This section describes our scenario design, survey development, and participant recruitment processes.


\subsection{Scenario Selection}
\label{subsec:scenario-selection}

Unlike back-end privacy issues such as data breaches, users' immediate sense of privacy hinges on design mechanisms. Drawing on deceptive design and privacy research across web (e.g.,~\cite{gray2021legal,habib2020scavenger,seyson2025exploring}), mobile (e.g.,~\cite{gunawan2021comparative,bosch2016tales,geronimo2020ui}), and VR platforms (e.g.,~\cite{trimananda2022ovrseen,hadan2024deceived,krauss2024makes}), and leveraging~\citet{gunawan2022redress}'s classification from literature and case law analyses, \revision{we iteratively clustered three design mechanism categories where deceptive patterns can manifest to influence users' privacy. Two researchers with expertise in usable privacy and deceptive design reviewed, discussed, and iteratively refined these categories based on the definitions and examples from the literature until reaching consensus.} These categories include: \textit{`Entry' requests} such as consent requests, privacy policies, user agreements, terms of service, and registration and subscription processes that users encounter when first accessing a service~\cite{gunawan2021comparative,gray2021legal,grassl2021dark}; \textit{user settings} such as privacy settings and specific permissions for data or sensor access that let users manage preference after initial consent~\cite{mildner2021ethical,bosch2016tales,geronimo2020ui}; and \textit{VR interactions} such as user behaviors within VR environments and interactions with virtual objects that can disclose personal information (see~\autoref{subsubsec:related-work-privacyuserinteraction}). These categories guided the VR design mechanisms and privacy-related interactions demonstrated in our scenarios.

\subsubsection{Privacy deceptive design patterns in commercial VR applications}
\label{subsubsec:gathering}

To ground participants' reflections in authentic VR contexts, guided by our three privacy mechanism categories, we drew on \revision{eight VR design mechanisms} from four commercial VR applications examined in prior study~\cite{hadan2024cscw}: VRChat\footnote{VRChart on Meta Quest.~\url{https://www.meta.com/experiences/vrchat/1856672347794301/}}, TRIPP\footnote{TRIPP: Meditation, Relaxation, Sleep on Meta Quest.~\url{https://www.meta.com/experiences/tripp-meditation-relaxation-sleep/2173576192720129/}}, Supernatural\footnote{Supernatural: Unreal Fitness on Meta Quest.~\url{https://www.meta.com/experiences/supernatural-unreal-fitness/1830168170427369/}}, and The Climb 2\footnote{The Climb 2 on Meta Quest.~\url{https://www.meta.com/experiences/the-climb-2/2617233878395214/}}. These applications were selected because they are consistently top-rated across VR app stores\footnote{See, for example, Meta Quest Store (\url{https://www.meta.com/experiences/view/1321443348416166/}) and SteamVR (\url{https://store.steampowered.com/vr/}) rankings, last accessed in February 2025.} and have previously been found to use a diverse range of deceptive patterns that threaten user privacy~\cite{hadan2024cscw}. The eight VR design mechanisms collectively demonstrate 11 deceptive patterns, all of which have been shown to manifest in mainstream VR applications and broader deceptive design and VR literature~\cite{hadan2024cscw,hadan2024computer,hadan2024deceived,krauss2024makes} (see~\autoref{tab:brief-scenarios}, middle column, for pattern descriptions). 

For ecological validity and methodological consistency, we adapted the media assets made available in the associated open-access dataset~\cite{hadan2024cscw} (e.g., screenshots and video recorded VR sessions\footnote{Open access materials available at: \textit{Privacy Threats from Deceptive Design in VR Games and Applications Dataset}, ~\cite{hadan2024cscw} 2025, Open Science Framework (OSF), \url{https://osf.io/axzve/}}) as our survey materials. To our knowledge, the dataset includes the materials from a top-tier, peer-reviewed examination~\cite{hadan2024cscw} of privacy-related deceptive design in commercial VR applications with a fully documented and replicable methodology, which provides us with a rigorous peer-reviewed foundation for scenario construction. We isolated single patterns in most scenarios to improve interpretability and reduce confounds, but S1-3 include multiple patterns to reflect the natural co-occurrence of deceptive design in real commercial VR applications under some contexts~\cite{hadan2024cscw} (see~\autoref{app-sec:scenarios}). 
We adapted 11 out of 14 deceptive patterns identified in previous work~\cite{hadan2024cscw} through a six-month, in-situ immersive first-person autoethnographic study in commercial VR applications.
We re-used VR recordings that captured in-the-wild deceptive patterns in existing VR applications. Using the three privacy mechanism categories as inclusion criteria, we excluded three privacy mechanisms in the sample that redirect users outside the VR environment (e.g., links to external webpages~\cite{hadan2024cscw}) or patterns that could not be evaluated without fabricating VR mechanisms (e.g., submissive acceptance where consent mechanisms were completely not presented~\cite{hadan2024cscw}). This resulted in our focused set of 11 patterns in eight privacy mechanisms, with nine associated with \textit{`Entry' requests}, one associated with  \textit{User settings}, and one associated with \textit{VR interactions} (see~\autoref{tab:brief-scenarios}). Each mechanism formed the basis for a survey scenario (see~\autoref{subsec:scenarios-construction}). We compiled 19 video recordings and 48 screenshots in Miro\footnote{Miro---the Virtual Workspace for Innovation.~\url{https://miro.com/}} as the supporting materials for our scenario construction. ~\autoref{tab:brief-scenarios} presents the eight scenarios, each including a description of possible user interactions within the given context, possible privacy-related decisions, associated privacy deceptive patterns, and the source VR application.

\subsection{Scenario Construction}
\label{subsec:scenarios-construction}

We constructed eight scenarios, each illustrating one of the eight selected VR mechanisms embedded with privacy deceptive patterns. We focused on VR mechanisms, instead of individual patterns, as deceptive patterns frequently co-occur in practice~\cite{Brignull2023book}. Each scenario included a brief description, two representative screenshots and a 10-second video showing an interaction with the mechanism, situating the scenarios in familiar VR application contexts to ground participants' reflections in their prior experience. The descriptions specified the mechanism's source application, its design elements, and possible user actions in the context. All descriptions followed a consistent structure to present information in a consistent level of detail. For each mechanism, participants first viewed an unmodified screenshot to establish reflection context. A second, annotated version of the screenshot was then provided to ensure participants could accurately locate the relevant design elements to answer questions about them. The annotations did not prescribe an interpretation but served to reduce ambiguity across diverse VR mechanisms, enabling participants to reflect on their own perceptions of potential manipulation.

We refined the descriptions, screenshots, and videos through multiple rounds of discussions during weekly research seminars at our institute with eight researchers in the field of HCI, VR, and game design who were not involved in our project. This process ensured that each scenario clearly captured the interaction flow and privacy decision moments that users must navigate when engaging with the mechanism. ~\autoref{tab:brief-scenarios} summarizes our scenarios, 
~\autoref{app-sec:scenarios} contains the full scenario materials, and short video clips are available in Supplementary Materials.


\begin{table*}[!t]
\caption{\revision{Overview of scenarios with incorporated deceptive design patterns~\cite{gray2024ontology,hadan2024cscw} (shown as colored tags) and privacy decisions. Participants viewed only the full scenario descriptions, screenshots, and video recordings provided in~\autoref{app-sec:scenarios}.}}
\label{tab:brief-scenarios}
\centering
\small
\renewcommand{\arraystretch}{1.3}
\resizebox{\textwidth}{!}{%
\begin{tabular}{@{}p{0.7\textwidth}p{0.4\textwidth}@{}}
\toprule
\textbf{Scenario (S) \& Deceptive Patterns~\cite{gray2024ontology,hadan2024cscw}} & \textbf{Privacy Decision \& User Data} \\ \midrule

\textbf{S1:} \textbf{VRChat} presents a \textbf{Notification} on first launch, asking users to grant or deny access to their photos, media, and files before proceeding. \par \vspace{2pt}
\tagFH~\tagBD~\tagPZ~\tagHI
& \textbf{Privacy Decision:} Grant device file access and save preference. \par \vspace{5pt}
\textbf{User Data}: Media files on VR device.
\\ \cdashlinelr{1-2}

\textbf{S2:} \textbf{TRIPP} presents a \textbf{Registration Form} requiring users to provide their name or email to unlock features, beyond the existing Meta account.\par \vspace{2pt} \tagFR~\tagPZ~\tagPNF
& \textbf{Privacy Decision:} Provide personal identifiable information. \par \vspace{5pt}
\textbf{User Data}: Full name and contact information.
\\ \cdashlinelr{1-2}

\textbf{S3:} \textbf{Supernatural} presents a \textbf{Subscription Form} with two plans and polished testimonials, requiring users to select and pay before accessing any features. \par \vspace{2pt}
\tagET~\tagFH~\tagPNF
& \textbf{Privacy Decision:} Disclose payment information. \par \vspace{5pt}
\textbf{User Data}: Payment information.
\\ \cdashlinelr{1-2}

\textbf{S4:} \textbf{The Climb 2} buries privacy policy under multiple layers of the \textbf{Settings Menu}, requiring navigation through sub-menus to locate it. \par \vspace{2pt}~\tagPM
& \textbf{Privacy Decision:} Seek and review privacy policy. \par \vspace{5pt}
\textbf{User Data}: No directly involved data.
\\ \cdashlinelr{1-2}

\textbf{S5:} \textbf{VRChat}  presents a \textbf{Consent Form} with a lengthy, complex privacy policy that users must scroll through and agree to before accessing the application. \par \vspace{2pt}~\tagCL
& \textbf{Privacy Decision:}  Agree to data collection terms. \par \vspace{5pt}
\textbf{User Data}: No directly involved data.
\\ \cdashlinelr{1-2}

\textbf{S6:} \textbf{The Climb 2} requests users to select their gender and skin color in the \textbf{Avatar Creation Mechanism}, with photorealistic hands rendered from a first-person perspective. \par \vspace{2pt}~\tagTQ
& \textbf{Privacy Decision:} Provide gender, skin color. \par \vspace{5pt}
\textbf{User Data}: Gender and skin color.
\\ \cdashlinelr{1-2}

\textbf{S7:} After denying file access in \textbf{VRChat}, users are repeatedly interrupted by a ``permissions error'' \textbf{Notification}, requiring them to press ``okay'' to proceed. \par \vspace{2pt}~\tagNG
& \textbf{Privacy Decision:} Grant permission to stop nagging. \par \vspace{5pt}
\textbf{User Data}: No directly involved data.
\\ \cdashlinelr{1-2}

\textbf{S8:} In \textbf{VRChat}, a 3D virtual \textbf{Signboard} guides users to locate and adjust their microphone settings. \par \vspace{2pt}~\tagNone
& \textbf{Privacy Decision:} Enable or disable microphone. \par \vspace{5pt}
\textbf{User Data}: Audio data.
\\
\bottomrule
\end{tabular}}
\end{table*}

\subsection{Survey Design}

To avoid priming participants, we advertised our study as an exploration of user experience and interactions in commercial VR applications. Recruitment materials, scenarios, and most survey questions deliberately avoided terms such as ``privacy,'' ``manipulation,'' and ``deception.'' Only at the end did we ask participants to reflect on privacy implications and general privacy concerns, after which we fully debriefed them on the study's purpose and our rationale for withholding this information. This approach was effective, as most participants reported positive attitudes towards the VR applications (see~\autoref{tab:brief-demographics}). 
\revision{Our full questionnaire is in Supplementary Materials.}



The survey began with an information sheet, consent form, and screening questionnaire. We recruited North American participants aged 18 or older with prior experience using one of four VR applications. VR device ownership was not required. Eligible participants then completed the main survey, which assessed their general VR familiarity and experience with a selected VR application, as these factors can affect their awareness, concerns, and susceptibility to privacy risks and manipulation~\cite{maloney2020anonymity,m2020towards,geronimo2020ui}. Specifically, we asked about their years of VR and application use (\varname{VR\_Experience} and \varname{App\_Experience}), purpose and frequency of use (\varname{VR\_Purpose} and \varname{App\_Use\_Frequency}), device brand (\varname{Brand}), professional experience in VR design, research, or development (\varname{VR\_ProExperience}), and satisfaction with the application, rated on a 5-point scale from ``1 star-poor experience'' to ``5 star-excellent experience'' (\varname{Star\_Rating}).


Participants were randomly assigned to a scenario based on a VR application they had previously used and reflected on how they would respond to the presented mechanism, drawing on their prior experiences~\cite{habib2022okay,bongard2021definitely}. To guide participants' attention toward privacy-related interactions, they answered a multiple-choice question on possible privacy decisions in the scenario (e.g., ``granting permissions,'' ``entering personal/payment information''). For each chosen option, they rated their perceived privacy impact of the selected decisions on a 5-point scale (``1-not at all'' to ``5-significantly'') and, if the rating was $\geq$3 (``moderately''), provided their explanations in an open-ended question.

Participants then identified specific design elements influencing their decisions by selecting them from an annotated screenshot. We provided a definition of design elements to participants using neutral language: \textit{``UI elements are the components of an interface that users interact with directly. Examples of UI elements include buttons, drop-down menus, sliders, text fields, checkboxes, toggles, and VR-specific features like realistic virtual objects or dynamic 3D environments.''} We used \textit{``UI elements''} as an umbrella term, defined broadly to also cover UX features, ensuring participants considered both without needing to distinguish them. Since perceived benefits of data sharing can influence user privacy concerns~\cite{naeini2017privacy}, participants were asked to rate the perceived benefits of each design element for users, the application's designer and publisher, and third-parties (e.g., marketing, legal, analytics) on a 5-point Likert scale (``1-strongly disagree'' to ``5-strongly agree''). We included two attention checks to ensure our data quality and participants' attention.


Finally, we assessed participants' general awareness of VR design's influence on privacy decisions (\varname{Influence}), perceived privacy harms (\varname{Harm}), and concerns about user manipulation (\varname{Concern}), adapting a measurement from~\citet{bongard2021definitely}. Six paired statements
captured participants' awareness and concerns about VR deceptive design from a general public (``people/others'') and personal (``myself/me'') perspectives on a 5-point scale \revision{(``1-strongly disagree'' to ``5-strongly agree'').} We adapted the original items by substituting ``websites'' with ``VR applications'' and linking VR design influence to personal data sharing. We also assessed participants' general online privacy concerns using the Internet Users’ Information Privacy Concerns Scale (IUIPC)~\cite{IUIPCscale}, which measures their perceived control over personal data (\varname{IUIPC\_Control}), awareness of data practices (\varname{IUIPC\_Awareness}), and concerns about data collection (\varname{IUIPC\_Collection}). Our survey concluded with demographic questions.


\subsection{Participant Recruitment and Demographics}
\label{subsec:demographics}

We piloted the survey with seven research group members, improved questions for clarity, and deployed the final survey on Prolific\footnote{Prolific | Easily collect high-quality data from real people.~\url{https://www.prolific.com/}} in December 2024. Participants received \revision{\$11.0 USD} for completion or \revision{\$1.1 USD} if screened out. From 576 complete responses (avg. completion time = 19.9 min), we excluded $n=3$ incoherent responses and $n=92$ responses showing AI-generated generic language that lacks concrete insights~\cite{hadan2024great}. 
The final sample included $N=481$ valid responses ($n\approx60$ per scenario), providing sufficient data for Wilcoxon signed-rank tests and logistic regressions without assuming interval-scale properties for ordinal data (power analysis: $\alpha\leq5\%$, effect size$=.50$, power=$.97$~\cite{faul2009statistical}).\footnote{Since G*Power does not support cumulative link mixed models (CLMMs) with random effects, we used logistic regression as a conservative approximation.}

\begin{table*}[!t]
\caption{Participant demographics ($N=481$), including gender, age, education, VR work experience, IUIPC scores, and awareness and concerns about the influence of VR design. Detailed demographics are in Supplementary Materials.}
\label{tab:brief-demographics}
\centering
\resizebox{\textwidth}{!}{%
\begin{tabular}{@{}lrllrllrllr@{}}
\cmidrule(r){1-2} \cmidrule(lr){4-5} \cmidrule(lr){7-8} \cmidrule(l){10-11}
\multicolumn{2}{c}{\textbf{Gender}} &  & \multicolumn{2}{c}{\textbf{Age}} &  & \multicolumn{2}{c}{\textbf{Education}} &  & \multicolumn{2}{c}{\textbf{Application Star Rating}} \\ \cmidrule(r){1-2} \cmidrule(lr){4-5} \cmidrule(lr){7-8} \cmidrule(l){10-11} 
Female & 216 (44.9\%) &  & Range & 18-64 &  & High school diploma, GED, or less & 67 (13.9\%) &  & \multicolumn{2}{l}{\textbf{\textit{\underline{VRChat}}}} \\
Male & 258 (53.6\%) &  & Mean (SD) & 33.37 (10.29) &  & Associates or technical degree & 47 (9.8\%) &  & Range & 1-5 \\ \cmidrule(lr){4-5}
Non-binary / third gender & 6 (1.2\%) &  & \multicolumn{2}{l}{} &  & Bachelor's degree or equivalent & 276 (57.4\%) &  & Mean (SD) & 4.00 (0.73) \\ \cmidrule(lr){4-5}
Prefer not to say & 1 (0.2\%) &  & \multicolumn{2}{c}{\textbf{Country of Origin}} &  & Graduate or professional degree & 91 (18.9\%) &  &  &  \\ \cmidrule(r){1-2} \cmidrule(lr){4-5} \cmidrule(lr){7-8}
\multicolumn{2}{l}{} &  & Canada & 95 (19.8\%) &  & \multicolumn{2}{l}{} &  & \multicolumn{2}{l}{\textbf{\textit{\underline{TRIPP}}}} \\ \cmidrule(r){1-2} \cmidrule(lr){7-8}
\multicolumn{2}{c}{\textbf{IUIPC Scores}} &  & USA & 386 (80.2\%) &  & \multicolumn{2}{c}{\textbf{VR Deceptive Design Awareness \& Concern}} &  & Range & 1-5 \\ \cmidrule(r){1-2} \cmidrule(lr){4-5} \cmidrule(lr){7-8}
\textbf{\textit{\underline{IUIPC\_Control}}} &  &  & \multicolumn{2}{l}{} &  & Range & \revision{1-5} &  & Mean (SD) & 3.88 (0.91) \\ \cmidrule(lr){4-5}
Range & 1-7 &  & \multicolumn{2}{c}{\textbf{Professional VR Work Experience}} &  & \textbf{\textit{\underline{People/Others}}}& &  &  &  \\ \cmidrule(lr){4-5}
Mean (SD) & 5.67 (0.97) &  & VR Developer/Engineer & 26 (5.4\%) &  & Mean (SD) & \revision{2.98 (0.94)} &  & \multicolumn{2}{l}{\textbf{\textit{\underline{Supernatural}}}} \\
\textbf{\textit{\underline{IUIPC\_Awareness}}} &  &  & VR Researcher & 58 (12.1\%) &  & Median & \revision{3} &  & Range & 1-5 \\
Range & 1-7 &  & VR Designer (e.g., UI, UX) & 40 (8.3\%) &  & Mode & \revision{4}  &  & Mean (SD) & 4.03 (0.89) \\
Mean (SD) & 6.14 (0.90) &  & VR Product Manager & 38 (7.9\%) &  & \textbf{\textit{\underline{Self/Me}}} & &  &  &  \\
\textbf{\textit{\underline{IUIPC\_Collection}}} &  &  & VR Data Analyst & 43 (8.9\%) &  & Mean (SD) & \revision{2.89 (0.93)} &  & \multicolumn{2}{l}{\textbf{\textit{\underline{The Climb 2}}}} \\
Range & 1-7 &  & Other & 5 (1.0\%) &  & Median & \revision{3} &  & Range & \revision{3-5 (of 1-5)} \\
Mean (SD) & 5.39 (1.33) &  & Never & 360 (74.8\%) &  &  Mode & \revision{4}  &  & Mean (SD) & 4.22 (0.65) \\ \cmidrule(r){1-2} \cmidrule(lr){4-5} \cmidrule(lr){7-8} \cmidrule(l){10-11} 
\end{tabular}%
}
\end{table*}


~\autoref{tab:brief-demographics} summarizes participant demographics. Our participants were 216 (44.9\%) female, 258 (53.6\%) male, 6 (1.0\%) non-binary, and 1 (.2\%) who preferred not to disclose, aged between 18 to 64 (M=33.4, SD=10.3). Most held a Bachelor's degree (57.4\%) or higher (Graduate or professional=18.9\%), had at least 1 year experience using VR devices (1 to 2 years=27.0\%, 2 to 5 years=36.6\%, More than 5 years=8.7\%), and 121 (25.2\%) had professional experience in VR research, development, or design. IUIPC scores showed generally high privacy concerns among participants (M$\geq5.39$), with \varname{IUIPC\_Awareness} being the highest (M=6.14, SD=0.90). Participants showed moderate awareness of how VR design influences data sharing decisions (M=3.27, SD=1.10), limited awareness of potential privacy harms (M=2.83, SD=1.18), and low concern about being manipulated when managing privacy in VR (M=2.71, SD=1.17).

\subsection{Data Analysis}

We report our quantitative data analysis with the results in~\autoref{sec:findings}. Open-ended responses were analyzed using inductive thematic analysis with two researchers~\cite{clarke2021thematic}. 
Upon familiarizing themselves with the data, each researcher independently coded 11\% ($n=50$) of the data, after which they met to discuss and merged developed codes and resolve disagreements. This process was repeated weekly for nine weeks, with an additional 11\% of data coded each week. The two researchers discussed, merged, and refined the codes and resolved disagreements iteratively every week until the codebook was finalized and all remaining data were coded. Using the finalized codebook, the researchers developed and refined the themes. An overview of our themes and codes \revision{is in Supplementary Materials.}

\subsection{Ethical Considerations}
\label{subsec:method-limitation}

We used a scenario-based online survey for three reasons. \textit{First,} privacy norms and human interpretations of choice architecture are highly context-dependent and subjective~\cite{nissenbaum2004privacy,mills2022personalized,sunstein2022sludge}. Scenarios (i.e., structured descriptions of real-world situations~\cite{finch1987vignette}) are widely adopted in privacy and deceptive design research to provide participants with contextual grounding that enables valid and interpretable responses~\cite{naeini2017privacy,hadan2024privacy,gray2021end,geronimo2020ui}. Although scenarios do not reproduce every nuance of lived experience, they offer a methodologically rigorous middle ground to allow researchers to probe situated judgments without exposing participants to uncontrolled or potentially harmful manipulations. \textit{Second,} unlike 2D interfaces, VR mechanisms are interactive, embodied, and immersive, making them difficult to evaluate through static screenshots or text alone~\cite{huart2004evaluation}. To increase ecological validity, we paired each scenario with a short video created by the researchers demonstrating via first-person perspective of interactions with the VR mechanism. This study design situates the scenario in relatable and familiar application contexts, helping participants connect their assessments and reflections to prior personal experience. Moreover, conducting a fully observational study in commercial VR applications would be logistically prohibitive; sessions are long, heterogeneous, and users may not naturally encounter all relevant privacy situations within a single playthrough. Through controlled video scenarios, we ensured that every participant engaged with the same set of mechanisms in a consistent way, while still grounding their responses in realistic VR contexts. \textit{Third,} scenario-based methods provide an ethical safeguard when investigating privacy-invasive deceptive designs. Directly exposing participants to manipulative VR mechanisms would risk violations of their privacy and autonomy. Our controlled scenarios simulated these situations and preserved participants' ability to reason about risks and choices while avoiding the harms of actual exposure. This aligns with established practices in privacy research, where simulated vignettes are routinely used to ethically study privacy sensitive decision-making~\cite{hadan2024privacy,naeini2017privacy}.

\section{Findings}
\label{sec:findings}
This section reports findings addressing our RQs (see~\autoref{sec:introduction}). Quantitative variables are in monospaced font (e.g.,~\varname{Influence\_self}), deceptive pattern names are \textit{italicized}, 
text directly extracted from VR mechanisms is in quotation marks (e.g., ``save preference''), and participant quotes are \textit{italicized} within quotation marks.

\renewcommand{\arraystretch}{1.2}
\begin{table*}[!t]
\centering
\caption{Overview of variables in our quantitative analyses. Demographic variables are reported separately in~\autoref{tab:brief-demographics}.}
\label{tab:variables}
\resizebox{\textwidth}{!}{%
\begin{tabular}{@{}ll@{}}
\toprule
\textbf{Variable} & \textbf{Description}$^a$ \\ \midrule
\varname{VR\_ProExperience} & Number of years that participants had prior experience with VR applications. \\
\varname{Star\_Rating} & Participants' overall evaluation of their experience with the VR   application. (5-excellent experience) \\
\varname{Influence\_self} & Participants' perceived influence of deceptive design on the their own   privacy decisions. (1-strongly disagree to 5-strongly agree) \\
\varname{Influence\_other} & Participants' perceived influence of deceptive design on the general   public's privacy decisions. (1-strongly disagree to 5-strongly agree) \\
\varname{Harm\_self} & Participants' perceived personal privacy harm from deceptive design   practices in VR. (1-strongly disagree to 5-strongly agree) \\
\varname{Harm\_other} & Participants' perceived privacy harm to the general public from deceptive   design practices in VR. (1-strongly disagree to 5-strongly agree) \\
\varname{Concern\_self} & Participants' personal concern about privacy impacts of deceptive design   in VR. (1-strongly disagree to 5-strongly agree) \\
\varname{Concern\_other} & Participants' concern about privacy impacts of deceptive design on the   general public. (1-strongly disagree to 5-strongly agree) \\
\varname{IUIPC\_Control} & Degree to which participants feel they can control their personal   information online. (1-strongly disagree to 7-strongly agree) \\
\varname{IUIPC\_Awareness} & Participants awareness of online platforms' general privacy practices.   (1-strongly disagree to 7-strongly agree) \\
\varname{IUIPC\_Collection} & Participants' concerns about the collection of personal data by general   online platforms. (1-strongly disagree to 7-strongly agree) \\
\varname{Benefit\_Me}$^b$ & Participants' perceived personal benefits from the VR interaction.   (1-strongly disagree to 5-strongly agree) \\
\varname{Benefit\_Dev}$^b$ & Participants' perceived benefits of the VR interaction to VR   applications' developers and publishers. (1-strongly disagree to 5-strongly agree) \\
\varname{Benefit\_Third-P}$^b$ & Participants' perceived benefits of the VR interaction to third parties.   (1-strongly disagree to 5-strongly agree) \\
\varname{Privacy\_Impact}$^c$ & Participants' perceived privacy impacts of their selected responses to the mechanism, drawing on prior experiences. (1-not at all to 5-significantly) \\ \bottomrule
\multicolumn{2}{l}{\begin{tabular}[c]{@{}l@{}}\textit{Note.} $a$. The detailed survey questions can be found in Supplementary Materials. \\ 
$b$. Conditional question shown only to participants who reported that one or more design elements in the VR mechanism influenced their interaction.\\
$c$. Conditional questions shown only to participants who reported that one or more decisions were made during their interaction with the VR mechanism. \\
\end{tabular}}
\end{tabular}%
}
\end{table*}

\renewcommand{\arraystretch}{1.0}


\subsection{RQ1: User Awareness and Concerns about Privacy Deceptive Design in VR}
\label{subsec:findings-RQ1}

We examined participants' general awareness and concern about VR deceptive design's privacy impacts from personal and public perspectives (\varname{Influence\_self} or \varname{\_other}, \varname{Harm\_self} or \varname{\_other}, and \varname{Concern\_self} or \varname{\_other}), overall VR application experiences (\varname{Star\_Rating}), perceived privacy impacts of decisions during interactions with each VR mechanism (\varname{Privacy\_Impact}), and perceived benefits to different stakeholders (\varname{Benefit\_Me}, \varname{Benefit\_Dev}, \\ \varname{Benefit\_Third-P}). These factors were found to influence users' perceived privacy risks and manipulation experiences in non-VR contexts~\cite{naeini2017privacy,gray2021end}. We discuss response distributions and their relationship to qualitative findings in~\autoref{subsec:findings-RQ3} below, with detailed descriptive statistics \revision{in Supplementary Materials.}

\subsubsection{Awareness, Harm Perception, and Concern}

Participants perceived VR design as more influential on others' privacy decisions (\varname{Influence\_other}: $median=1$) than their own (\varname{Influence\_self}: $median=0$). Similarly, they expressed greater uncertainty about manipulation risks for others (\varname{Concern\_other}: $median=-1$) than for themselves (\varname{Concern\_self}: $median=0$). Wilcoxon signed-rank comparisons~\cite{woolson2005wilcoxon} with Bonferroni correction~\cite{chen2017general} confirmed both differences as significant (\varname{Influence}: $P=.001$, \varname{Concern}: $P=.004$). However, no significant difference was found in participants' perceived harm, with both perspectives remaining uncertain (General: $median=0$, Personal: $median=0$, \varname{Harm}: $P=.94$). 



\subsubsection{General positive experience with VR applications}


As shown in~\autoref{tab:brief-demographics}, participants reported generally positive VR application experiences (\varname{Star\_Rating}: M$\geq$3.88, SD$\leq$0.91). Most ($n=372, 77\%$) participants mentioned soothing atmosphere and high-quality audio and visuals, with a small number ($n=15, 3\%$) emphasized real-world well-being benefits. These align with our qualitative results in~\autoref{subsubsec:RQ3-realistic}, where lifelike simulations and calming music contributed to participants' willingness to disclose personal information when encountering deceptive patterns. Participants' negative feedback primarily focused on technical and accessibility issues or incompatible social dynamics ($n=103, 21\%$), consistent with qualitative reports of participants mistaking S7's (\pattern{nagging}) for a technical flaw (see~\autoref{subsubsec:RQ3-none}). A few participants criticized the monetization, with 7 ($1.4\%$) complained about unfair pricing in Supernatural (S3), and 2 ($<1\%$) described TRIPP's (S2) paid experience as \textit{``a bit clunky and unrefined''} compared to the free trial. Qualitative data from~\autoref{subsubsec:RQ3-realistic} further revealed participants' privacy concerns about S2's lack of a free trial and S3's upfront payment requirement. These frustrations indicate a potential \pattern{Bait and Switch}~\cite{gray2024ontology}, where user expectations are subverted and payment information is compromised. Our codebook is included \revision{in Supplementary Materials.}


\subsubsection{Perceived privacy impacts and benefits from interacting with the VR mechanisms}

\begin{figure*}[!t]
    \centering
    \includegraphics[width=0.9\textwidth]{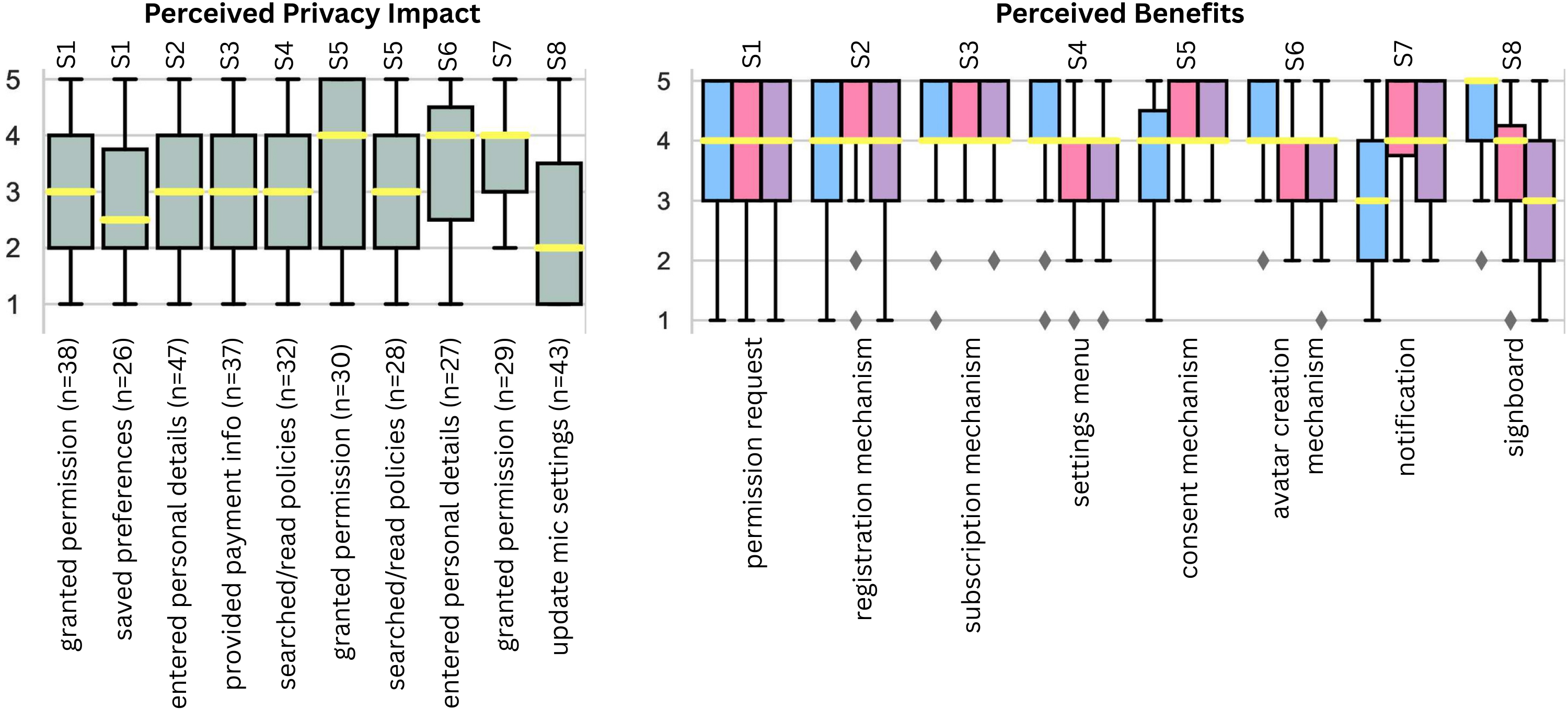} 
    \vspace{-2mm}
    \caption{S=Scenario. Participants' ($N=481$) responses regarding: \textbf{Left:} (conditional question) perceived privacy impacts of their selected responses to the presented mechanism in each VR scenario, drawing on prior experiences with the application (1-not at all to 5-significantly); and \textbf{Right:} perceived benefit of each design mechanism for themselves (user~\textcolor[HTML]{85C4FF}{\faSquare}), the VR application's developer and publisher (Dev \& Pub~\textcolor[HTML]{FF85B1}{\faSquare}) and third-parties stakeholders (e.g., marketing, legal services, analytical vendors) (Third-P.~\textcolor[HTML]{BE9FD1}{\faSquare}). (1-strongly disagree to 5-strongly agree). Higher scores indicate greater perceived privacy impact or perceived benefit, respectively. \revision{Overall, participants perceived the highest privacy impact from S5's consent mechanism and S7's permission notification, rating both as more beneficial for developers and third parties than for themselves.}}
    \Description{Participants' boxplots}
    \label{fig:barcharts}
\end{figure*}

As ~\autoref{fig:barcharts} shows, participants perceived higher \varname{Privacy\_Impact} from ``entering personal details'' in S6's avatar creation ($n=27, median=4$), ``granting permission'' in S7's repeated permission notifications ($n=29, median=4$) and in S5's consent mechanism ($n=30, median=4$). The mechanism in S6 was perceived as more beneficial for users than for developers or third parties (e.g., marketing, legal services, analytical vendors), echoing participants' qualitative responses in~\autoref{subsubsec:RQ3-realistic} where avatar creation was valued for supporting self-expression and realism, with privacy concerns being mitigated by trust and enjoyment. Conversely, mechanisms in S7 and S5 were seen as less beneficial for users ($median=3$) than for other parties ($median=4$), with qualitative responses in~\autoref{subsec:findings-RQ3} frequently describing these mechanisms as technical bugs or non-VR-style policies that interrupted the experience.

In contrast, participants perceived low \varname{Privacy\_Impact} for ``saving preference'' in S1's permission request ($n=26, median=2.5$), echoing~\autoref{subsubsec:RQ3-none} where participants viewed the pre-checked box (\pattern{bad default}~\cite{gray2024ontology}) as a convenience. Participants also rated the ``updating microphone settings'' in S8 as having the lowest \varname{Privacy\_Impact} ($n=43, median=2$) and the most beneficial for users ($median=5$), which aligns with its role as a no-pattern condition that merely offers guidance without directing user behavior. Other mechanisms were generally rated as having moderate privacy impact ($median=3$), with no notable differences in perceived benefits across the three parties ($median=4$).
\subsection{RQ2: Factors Contributing to Perceived Privacy Impacts and Benefits}
\label{subsec:findings-RQ2}

We further conducted Cumulative Link Mixed Model (CLMM) regression\footnote{We used the Ordinal R-package (\url{https://cran.r-project.org/web/packages/ordinal/}) for modeling participants' perceived privacy impacts and benefits from interacting with the VR design mechanisms.} to examine how participants' perceived privacy impacts and benefits from interacting with VR design mechanisms 
related to the presence of deceptive patterns (\varname{Scenario\_Type}), demographics, general online privacy concerns (IUIPC scores~\cite{IUIPCscale}), general awareness and concern of the privacy impacts of VR manipulation from both personal and general public's perspectives (\varname{Influence\_self} or \varname{\_others}, \varname{Harm\_self} or \varname{\_others}, and \varname{Concern} \varname{\_self} or \varname{\_others}), and overall VR applications experience (\varname{Star\_} \varname{Rating}). CLMM regression is well suited for repeated-measures designs with ordinal dependent variables, as in our study, where participants' perceived benefits were assessed multiple times for specific VR design elements, and their perceived privacy impacts were assessed multiple times for each of reported interactions~\cite{christensen2019tutorial}.
Specifically, we conducted four separate multivariate CLMM regressions, each using participants' responses to \varname{Privacy\_Impact}, \varname{Benefit\_Me}, \varname{Benefit\_Dev}, or \varname{Benefit\_Third-P} as the dependent variable (DV) and using other variables as the predictors. 

We obtained the best-fit model for each regression using backward stepwise elimination, iteratively removing the predictor that most reduced Akaike Information Criterion (AIC) from a full model until reaching global minimum AIC~\cite{kadane2004methods}. \revision{The resulting CLMM models are in~\autoref{app-sec:statistics}.} 
\revision{We highlight statistically significant predictors in this section.}



As shown in~\autoref{tab:CLMM}, \varname{Scenario\_Type} was negatively associated with \varname{Benefit\_Me} ($P<.001$) but positively associated with \varname{Benefit\_Dev} ($P=.04$), and \varname{Benefit\_Third-P} ($P<.001$). These findings suggest that participants viewed VR mechanisms involving deceptive design as more beneficial to developers and third parties than users like themselves. Both \varname{VR\_ProExperience} and \varname{Star\_Rating} were positively associated with \varname{Benefit\_Me} ($P<.001$ and $P<.001$), suggesting that participants with professional VR experience and positive application experiences were more likely to recognize the VR mechanisms' user benefits. 
In addition, \varname{IUIPC\_Control} was negatively associated with \varname{Benefit\_Me} ($P=.04$). This indicates that participants with stronger concerns about personal information control were less likely to view VR mechanisms as beneficial to users, \revision{suggesting they may have recognized these mechanisms as limiting their ability to make informed privacy decisions.} \varname{IUIPC\_Awareness} was positively associated with \varname{Benefit\_Me} ($P<.001$) and \varname{Benefit\_Third-P} ($P<.001$), indicating that participants with greater awareness of online privacy risks were more likely to perceive VR mechanisms as beneficial for both users like themselves and third-party stakeholders.

Moreover, we found statistically significant associations between participants' perception of VR design and their perceived benefits and privacy impact. Their concern about manipulation harms to others (\varname{Concern\_others}) was negatively associated with perceived benefits for users like themselves (\varname{Benefit\_Me}: $P=.03$), and their perceived manipulative influence on users like themselves (\varname{Influence\_self}) was positively associated with their perceived privacy impacts (\varname{Privacy\_Impact}: $P<.001$). Thus, participants who expressed greater concern about harms to others were less likely to view VR mechanisms as beneficial to users like themselves, and those who felt more susceptible to manipulative influence were more likely to recognize privacy consequences of their interactions.

\subsection{RQ3: VR-Specific Properties Shaping User Responses to Deceptive Design}
\label{subsec:findings-RQ3}

Drawing on participants' prior experience with the VR application, we asked how they would react to the VR mechanisms to understand their interpretations and responses to privacy deceptive patterns. To make clear the interplay between VR-specific properties and these patterns, we organized participants' responses based on how VR properties influenced the way in which deceptive patterns impact participants' privacy understanding and decisions across the scenarios. 
For clarity, we report the percentage of participants (e.g., 21\%) who mentioned a theme in a scenario, calculated based on the total participant number in the scenario. We denote the source of participant quotes in the parentheses (e.g., S1-P42).

\subsubsection{Realistic simulation and multi-sensory feedback.}
\label{subsubsec:RQ3-realistic}
VR systems enable realistic sensorimotor interactions by delivering high-fidelity visual, motion, audio, and haptic feedback that synchronizes users' movements with virtual objects in real time~\cite{skarbez2021revisiting,chan2011dance}. Virtual objects can be anchored to users' body, which produces a strong sense of presence in virtual worlds~\cite{slater2009place,skarbez2021revisiting}. The multi-sensory feedback allows users to feel virtual objects~\cite{salisbury1997phantom}, hear realistic sound propagation from different directions~\cite{savioja2015overview,liu2020sound}, and experience consistent spatial perception as visual and auditory cues align with head orientation~\cite{Slater1997framework}. Some VR environments also incorporate a virtual body that anchors user perception to a first-person perspective through its eyes and position~\cite{ellis1995nature,Slater1997framework}. Our findings show that participants were susceptible to privacy deceptive patterns that framed as enabling immersion or preventing disruption, and were willing to trade personal information for continued realistic and multi-sensory VR experiences.


\paragraph{\textbf{Forced Registration and Privacy Zuckering.}}
In S2, users encountered a mandatory registration mechanism that requires their full names and email address (\pattern{forced registration}~\cite{gray2024ontology}), with no free trial being offered beforehand (\pattern{privacy zuckering}~\cite{gray2024ontology}, see~\autoref{app-sec:scenarios}). This mechanism appears within an animated, space-themed environment demonstrating the application's \textit{``calming visuals and relaxing music''} (S2-P104) that helped participants to \textit{``clear my mind and soul''} (S2-P80). Eleven (19\%) reflected on this dynamic virtual background as the primary reason for their registration, with 3 (5\%) described it as engaging and inviting, and 7 (12\%) noted it sparked their curiosity by offering \textit{``a tiny taste''} of what the application might provide. Four (7\%) participants added that the immersive background made them feel positively about the application and made the request for full names feel acceptable. 

\begin{quot}{I'm willing to give my details because it seemed calming and peaceful.}{S2-P85}
\end{quot}

Notably, the VR application does not offer a formal free trial before registration, leaving users to base their expectations solely on the virtual background. As a result, 13 (22\%) participants reported developing coping strategies, 
such as using fake or abbreviated names or secondary email accounts, though they recognized that such strategies would be ineffective in applications requiring verification (e.g., confirmation codes). In addition, 23 (39\%) participant reflected on feeling irritated by being required to provide their full name. 
Four (7\%) described the registration as \textit{``buzzkill''}, 14 (21\%) raised concerns about potential misuse or sale of their personal data for advertising, and six (10\%) viewed the collection of full name as unnecessary. Worryingly, five (8\%) participants stated they felt they had to provide their information to move forward, 
even though they would have preferred not to share it.

\begin{quot}{I really dislike when an app asks for login information before even seeing what it's about.}{S2-P70}
\end{quot}

In contrast, 35 ($59\%$) participants mentioned providing personal information in the registration mechanism without hesitation, 
with 10 (17\%) trusted the publisher to protect their data. Many viewed the registration mechanism's design as standard ($n=27, 46\%$) or typical in non-VR applications ($n=20, 34\%$), which lowered their concerns. 

\paragraph{\textbf{Endorsements and Testimonials.}}
In S3, users encountered a subscription mechanism embedded in a mountain-themed VR environment, where the \textit{``high-quality visuals and audio''} were frequently described as evoking a sense of presence in a \textit{``breathtaking''} lifelike world (S3-P151). The mechanism also incorporated testimonials to motivate user subscriptions (\pattern{endorsements and testimonials}~\cite{gray2024ontology}, see~\autoref{app-sec:scenarios}), which requires users to disclose their payment information for the subscriptions. Nineteen (31\%) participants described this realistic virtual environment as providing \textit{``a glimpse of what I could experience,''} which made them \textit{``excited to sign up''} (S3-P79). 
Twenty-nine ($48\%$) chose the longer free trial offered with the annual subscription so they could explore the the high-quality graphics more thoroughly. Forty (66\%) participants even wished for even longer trial periods to better evaluate the experience. 

\begin{quot}{When I put that headset on and got in I saw that realistic landscape and was already hooked.}{S3-P165}
\end{quot}

In addition, 11 (18\%) participants stated that testimonials, combined with the \textit{``exciting''} and \textit{``beautiful''} landscape, made them confident in the subscription's value. 
On the other hand, 16 (26\%) expressed concerns and reluctance, 
with 6 ($10\%$) questioned the authenticity of the testimonials, describing them as \textit{``too polished''} and sought feedback from other users. 
Moreover, 7 (11\%) participants found the requirement to enter payment information for a free trial problematic, and perceived this subscription mechanism as relying on users to \textit{``forget to cancel before the trial ends''} (S3-P116).

\begin{quot}{If it's a free trial and then asks for your payment information, it doesn't feel like a free trial.}{S3-P119}
\end{quot}

In contrast, 34 (56\%) participants considered the overall subscription design standard for such applications, 
while only two ($3\%$) participants reflected on being swayed by positive recommendations from friends or family. 





\paragraph{\textbf{Trick Questions.}}
In S6, users encountered an avatar creation mechanism that asked them to ``select your gender and skin color'' (\pattern{trick questions}~\cite{gray2024ontology}), with changes immediately reflected on realistic virtual hands shown in front of them (see~\autoref{app-sec:scenarios}). Thirty-five (57\%) participants reported that the realistic virtual hands were important in their avatar creation, with 16 (26\%) considered them as facilitating accurate avatar customization, and 13 (21\%) viewed them as enabling personal expression. 
Twenty-one (34\%) participants mentioned that directly seeing changes in skin tone and gender, especially when paired with the \textit{``realistic visual and audio''} effects (S6-P304), made them feel the avatar's body as an extension of their own and motivated them to create avatars that \textit{``resemble my appearance in real life''} (S6-P284). These participants explained that this resemblance made their VR experience feel more authentic and personal. 
Among these participants, 12 (20\%) specifically reported selecting their actual skin tone and 9 (15\%) provided their physical gender. Furthermore, 55 (90\%) participants stated they \textit{``felt assured from the text [instruction]''} that they should \textit{``chose the skin color and gender that fit me''} (S6-P288), which in turn led them to align their avatars with their physical identity.

\begin{quot}{Because I'm African American. I preferred some ethnic or color to my hands.}{S6-P317}
\end{quot}


Worryingly, privacy was barely a concern of participants, with only 1 (2\%) participant mentioned they \textit{``checked privacy settings to make sure nothing's shared without my consent''} (S6-P316). 
In contrast, 39 ($64\%$) participants expressed trust in the application and positive impressions of the mechanism's customizability, which lessened their concerns. 
Seven (11\%) participants viewed selecting gender and skin color as a normal and expected feature of such applications, 
and two (3\%) participants further noted that \textit{``if the skin color and the gender of the hands were different from the my hands, the experience would be less immersive''} (S6-P312). 
Eleven (18\%) participants even expressed a desire for more diverse and realistic skin tone options to better reflect their actual identity in the physical world. 

\subsubsection{Device ergonomic challenges.}
\label{subsubsec:RQ3-ergonomic}

While VR offers engaging experiences, the weight, size, and design of existing head-mounted displays (HMDs) is a known source for ergonomic discomfort~\cite{hirzle2021critical}. Studies have found users' feelings of VR sickness, digital eye strain, and neck and shoulder issues~\cite{chang2020virtual,hirzle2021critical,hirzle2022understanding}. Our findings indicate that these discomforts lowered participants' patience during interactions, leading to fast and careless decisions. 

\paragraph{\textbf{Privacy Maze.}}
In S4, the full privacy policy was buried several layers deep in the settings menu (\pattern{privacy maze}~\cite{gray2024ontology}), requiring extra effort from those who wished to locate the VR application's data practices (see~\autoref{app-sec:scenarios}). Three (5\%) participants perceived this menu structure as \textit{``needlessly complicated''} (S4-P192), and five (9\%) reported difficulty finding the privacy information. 
One (2\%) participant specifically noted that \textit{``all the colors and shapes [were] throwing me off,''} which made it harder for them to navigate the menus inside VR (S4-P198). Participants' difficulties with menu navigation were further exacerbated by their physical discomfort, as six (10\%) reported they experienced motion sickness and neck pain from the application's head-movements.

\begin{quot}{I don't like that it makes you crane your neck. That gets uncomfortable after a while.}{S4-P211}
\end{quot}

Twenty-five (43\%) participants mentioned they turned to external sources to find privacy information, 
14 (24\%) relied on explanations from friends already using the application, and nine (16\%) wished for simpler navigation of privacy information. In contrast, 17 (29\%) participants appreciated the layered design because it kept the interface simple and clear, 
and 11 (19\%) viewed it as a standard design layout in other applications. 

\begin{quot}{It took forever to figure out where the privacy policy was because it was hard to read on Quest 2.}{S4-P215}
\end{quot}

\paragraph{\textbf{Complex \& Lengthy Language.}}
In S5, users encountered a long, complex privacy policy upon first entry (\pattern{complex \& lengthy language}~\cite{gray2024ontology}) and were required to consent to data collection within the same mechanism (see~\autoref{app-sec:scenarios}). Thirty-eight (59\%) participants appreciated receiving privacy information upfront, 
and 17 (27\%) valued the transparency of privacy information. 
At the same time, 23 participants (36\%) reported little motivation to fully read the policy, 
with 21 (34\%) admitted they agreed quickly without reading, and 8 (13\%) referred to it as a \textit{``wall of text... causes my eyes to look for the escape route... the agree button''} (S5-P248). Among these participants, 4 (6\%) also mentioned that the discomfort from the VR headset's weight and their eagerness to begin the VR experience lowered their patience for reading.

\begin{quot}{My ability to have patience when I am wearing a few pounds of equipment on my head is not great.}{S5-P351}
\end{quot}

Twenty-seven (42\%) participants viewed lengthy policies as standard for consent mechanisms. 
Still, 14 (22\%) concerned about vague data collection explanation, especially since it involved eye tracking. 
Seven (11\%) participants also expressed frustration with the length and complexity, referring to reading it as tedious and time-consuming. As a result, coping strategies were developed by 16 (25\%) participants, 
such as seeking simpler explanations and concise summaries online. One (3\%) participant also suggested displaying the privacy policy on a secondary device reduce the headset's ergonomic burden. On the contrary, six (9\%) participants believed they must \textit{``agree to it in order to move forward''} (S5-P227). 

\begin{quot}{Launch a window before you put your headset on so people will have the patience to look through it.}{S5-P273}
\end{quot}

\subsubsection{Non-deceptive scenario and the persistent 2D deceptive patterns in VR environments}
\label{subsubsec:RQ3-none}

Finally, the majority of participants ($n=54, 95\%$) perceived the signboard in S8 as providing straightforward guidance that \textit{``gives me full control on the management of my audio and who can hear me''} (S8-P372). We also found no notable effects of VR-specific properties on participants' responses to the \pattern{hidden information}, \pattern{false hierarchy/visual prominence}, \pattern{bad default}, and \pattern{nagging}. In S1, the application requested access to users' media files without adequate explanation and before related functions were used, with the ``allow'' button highlighted in vibrant blue and the ``save my preference'' checkbox pre-checked by default (see~\autoref{app-sec:scenarios}). In S2, the registration mechanism emphasized the ``next'' button in vibrant green, and in S3 the subscription mechanism highlighted the ``subscribe'' and ``annual plan'' options in the same color. These design choices made the highlighted options visually salient and, in S2 and S3, conveyed the \textit{``universal meaning of go,''} which some participants described as \textit{``a good touch to add to the experience''} (S3-P163). In S7, after declining the initial media file permission request in S1, users encounter repeated reminder notifications that participants described as \textit{``very irritating''} and \textit{``annoying''} but attributed as technical malfunction rather than manipulative design. Overall, participants perceived these patterns in ways similarly to prior non-VR research (e.g.,~\cite{gunawan2021comparative,gunawan2022redress,gray2021end}).

\section{Discussion}
\label{sec:discussion}



Our study extends prior work by exploring how users' perceptions of VR affordances, shaped by their own VR experiences, inform their responses to privacy deceptive design. 
Participants' desire for explicit consent opportunities and clear data collection explanations echoes existing findings~\cite{hadan2024privacy,harborth2021investigating}, yet our results show that VR's immersive nature and sensorimotor properties can further reduce impair users' privacy decision and heighten their deceptive design susceptibility (see~\autoref{subsec:findings-RQ3}). 
Since many current VR privacy interfaces still rely on 2D overlays, some patterns of influence resemble those observed in non-VR contexts (e.g.~\cite{bosch2016tales,gray2021end}) or theorized about VR (e.g.~\cite{hadan2024deceived,krauss2024makes}). For example, participants' familiarity with deceptive patterns in non-VR contexts often lowered their concerns about encountering similar manipulations in immersive settings. At the same time, participants suggested VR-specific vulnerabilities, such as bodily engagement and the physical discomfort of wearing cumbersome head-mounted headsets, may amplify the impact of deceptive design, as these factors can reduce user patience, increase decision fatigue. In this section, we discuss the implications of these findings for deceptive design research in immersive environments and outline actionable directions for future work.

\subsection{Embodied Susceptibility in VR and the Potential Role of Physical Discomfort}

Research on embodiment and immersion suggests that immersion and sensorimotor contingencies foster strong feelings of ``being there'' and make the depicted situations feel real~\cite{kilteni2012presence,slater2009place}. Participants noted that perceiving their avatar as an extension of their body could be exploited for manipulation. This suggestes that presence and plausibility illusion may reduce users' scrutiny of privacy choices as the virtual world feels more bodily and perceptually real. These cases may reflect participant sentiments of literature's hypothesis that realistic VR simulations amplify deception~\cite{hadan2024deceived,ramirez2021what,adams2018ethics}.

Although prior work suggests that VR's full immersion can limit users' capability to seek external privacy information when encountering convoluted VR privacy policies~\cite{hadan2024cscw,ramirez2021what}, our participants did not report influences commonly attributed to these physical-virtual barriers~\cite{mann2018all,mann2023fundamentals,krauss2024makes}, nor did they describe  experiences resembling emotional, hyper-personalized, and hardware-driven manipulation~\cite{krauss2024makes,hadan2024deceived,tang2025dark,liu2025flytrap}. We suspect this may related to current VR hardware's limited support for advanced interaction modalities.
However, our participants highlighted ergonomic issues (e.g., neck and eye discomfort) from prior VR use, noting that such discomfort can exacerbate the impact of deceptive design by reducing user patience in decision-making and prompting hasty or less-careful decisions, particularly when encountering \pattern{privacy maze} and \pattern{complex \& lengthy language} that have already made privacy information difficult to navigate~\cite{jensen2004privacy,linden2018privacy,adjerid2013sleights}. This suggests that users' privacy vulnerabilities may extend beyond informational or cognitive domains and may arise from physical strain. 
Existing privacy and deceptive design research has primarily focused on interface layout, user flow, and human cognition (e.g.,~\cite{gray2025getting,schlembach2021forced,su2022perception}). The potential for bodily and physiological strain to steer users' privacy decisions suggests that ergonomics could be examined as an additional design dimension.

\textit{Implications for Future Research:} Many leading VR persuasion theories assume fully immersive experiences and strong bodily engagement~\cite{slater2009place,kilteni2012presence}, but most current commercial VR systems only partially provide these properties. Looking forward, we propose that, beyond software-based VR properties (e.g.,~\cite{krauss2024makes,hadan2024deceived}), future research on VR deceptive design can proactively consider the possibility of how emerging VR hardware capabilities (e.g., electrical muscle stimulation~\cite{liu2025flytrap}, ergonomics, and physiological sensing~\cite{miller2020personal,buck2022security}) may introduce further new forms of manipulations. 

We further suggest that usability evaluation frameworks for VR privacy choice mechanisms consider incorporating ergonomic criteria, alongside traditional dimensions such as user comprehension and sentiment~\cite{habib2022evaluating,feng2021designspace}, to more holistically address users' privacy decision-making challenges in immersive environments and ensure VR designs support users' cognitive and physical needs in privacy decision-making. Prior work has shown that ergonomics affect users' attention, eye-tracking accuracy, head-hand coordination, and motion sickness~\cite{Mimnaugh2023virtual,zabihhosseinian2019neck,hirzle2021critical,kilteni2012presence}. Our participants mentioned that physical discomfort can reduce their willingness to engage with privacy mechanisms, potentially making certain manipulative tactics more influential (see~\autoref{subsec:findings-RQ3}). For example, literature shows that ergonomic issues can impair eye-hand coordination~\cite{zabihhosseinian2019neck} and lead to motion sickness~\cite{Mimnaugh2023virtual}. These conditions may create situations in which users are more likely to misclick or feel motivated to leave uncomfortable screens quickly, which in turn could lead to acceptance of invasive data requests.
Therefore, we propose the term \textbf{Ergonomic Susceptibility} as a potential new category of deceptive design that is worth considering in existing taxonomies~\cite{hadan2024deceived,ramirez2021what}. This category captures situations in which known patterns, such as those in~\autoref{subsubsec:RQ3-ergonomic}, become more effective due to hardware-induced ergonomic issues. We encourage future research to move beyond neck and eye strain reported by our participants and employ dedicated ergonomic measures (e.g.,~\cite{hirzle2021critical}) to systematically assess how a broader range of issues (e.g., shoulder pain, motion sickness, and facial pressures~\cite{chang2020virtual,hirzle2022understanding}) may create new vulnerabilities for deceptive design can exploit. Such research may also inform VR design recommendations and regulations that protect users from hardware-induced privacy risks.



\subsection{Immersion Normalizes Harmful VR Data Practices}
\label{subsec:immersive-contexts}


While many VR deceptive patterns resemble those in web and mobile contexts~\cite{hadan2024cscw,gunawan2021comparative}, participants' responses indicated the possibility that immersion may intensify their impact. Our participants frequently reported that they might be compelled to disclose personally identifiable and demographic data when requests were framed to sustain the sense of presence, even when such data is functionally unnecessary (see~\autoref{subsubsec:RQ3-realistic}). Regression results further showed that users' perceived benefits of VR mechanisms positively correlated with overall experience, suggesting that immediate benefits from VR affordances may skew the traditional privacy calculus~\cite{li2012theories,dinev2006extended}, especially since the privacy harms are often delayed into the future (i.e., hyperbolic time discounting~\cite{acquisti2017nudges}). 

Our research provides early evidence on how VR immersive properties that enable embodied and realistic experiences, combined with \textit{trust cues} that signal system trustworthiness, may diminish users' privacy concerns and normalize harmful data practices as routine aspects of VR experiences. For example, our participants' described how the possible intense feeling of self-location and agency inside virtual avatars could motivate their willingness to share physical identity details for coherent body representation, and their trust in the publisher might lead them to believe that their data would be protected (see~\autoref{subsec:findings-RQ3}). Research shows that users often experience logos and ads that commercial VR applications use to foster brand trust as enjoyable and engaging~\cite{mhaidli2025intriguing}, and that brand trust in non-VR contexts can be cultivated through rules, narratives, and mechanics that communicate brand values~\cite{bogost2010persuasive}. 
In VR, users are placed within realistic, sensory-rich, and fully immersive environments where narrative and mechanics are often experienced from a first-person perspective, fostering the sense of embodiment and empathy to the rendered experiences~\cite{hadan2024deceived,kilteni2012presence,schlembach2021forced}. 

\textit{Implications for Future Research:} We recommend future research investigate how branded virtual environments, combined with immersive properties like sensory richness and realistic simulations, influence users' responses to deceptive design and data disclosure requests. For instance, when such requests are embedded into storyline progression and dialogues or non-verbal interactions with non-player characters~\cite{frommel2012gathering,king2023investigating}. Further, as VR platforms integrate modalities such as haptics, olfactory cues, and full-body tracking, we also suggest future researchers to examine how advanced modalities may relate to users' perceptions of privacy risks and their susceptibility to deceptive design across diverse user groups. 


\subsection{Rethinking Solutions for VR-Specific Deceptive Design Challenges}


While our participants struggled to identify the exact design elements that could cause discomfort, they implicitly recognized the possibility that certain VR mechanisms might disrupt their experience and threaten their privacy (see~\autoref{subsec:findings-RQ3}). From VR designer and publisher's perspective, such implicit awareness can diminish user adoption and undermine applications' long-term sustainability~\cite{hadan2024ow2,vitell2003consumer}. However, existing ethical design efforts have mainly been developed for 2D web and mobile contexts~\cite{gray2024building,Chivukula2024surveying,chivukula2025universal}, where interactions are screen-based and mediated through conventional input devices (e.g., keyboards, mice, controllers). In contrast, VR design extends across a 360-degree environment rendered by users' visual field~\cite{skarbez2021revisiting,steinicke2016being} and relies on multimodal sensing of users' movement, visual attention, and physiological state~\cite{miller2020personal,pfeuffer2019behavioural,buck2022security}, creating privacy risks and manipulative opportunities beyond the scope of established 2D ethical design solutions. Despite existing regulatory frameworks for protecting user privacy and autonomy (e.g.,~\cite{California2020CPRA,EUDSA2022}), many deceptive patterns fall into gray areas with unclear enforcement~\cite{tran2025dark,noyb2021where,sanchezrola2019can}. However, our findings show that many deceptive patterns from non-VR contexts may remain effective in VR, and many participants described a sense of privacy resignation, possibly perceiving these patterns as \textit{``standard''} and inevitable across VR and non-VR environments. This resignation is especially concerning in VR, with VR's passive sensing and inferences can introduce unexpected privacy risks~\cite{hadan2024privacy} and immersive properties further lower users' privacy concerns and their resistance to data disclosure requests (as discussed in~\autoref{subsec:immersive-contexts}).

\textit{Implications for Future Research:}
As VR devices and applications continue to evolve, we encourage future VR researchers, industry practitioners, and policymakers to expand current ethical design approaches and regulatory frameworks to immersive contexts. A possible direction could be examining how trust is communicated in VR to avoid inadvertently concealing manipulation, and how VR's immersion and ergonomics might be leveraged to support user agency and informed decisions. For instance, rather than relying on disruptive text-based pop-ups that break immersion, designers might embed a companion character or a brief tactile vibration in controllers to communicate privacy-sensitive decisions. 
Our findings also point to the possibility that ergonomic discomforts may become instrument of manipulation by reducing users' likelihood of thoroughly considering data disclosure requests. While future improvements in VR hardware will likely reduce these issues over time, we believe that developers should still be mindful of how immersive, multimodal, and ergonomic properties might introduce novel vulnerabilities for user privacy and autonomy. 

\subsection{Reconsidering Consent and Transparency Mechanisms in VR}



Many participants proposed coping strategies in response to data practices they found uncomfortable, such as fabricating personal information or consulting external sources for guidance (see~\autoref{subsec:findings-RQ3}). Participants also expressed a desire for making complex or critical privacy decisions outside immersive sessions, such as through a companion application or desktop launcher. These findings indicate that the consent and privacy communication mechanisms implemented in commercial VR applications, directly borrowed from 2D computer and mobile contexts, may not fully support users' privacy needs in immersive environments. However, moving privacy decisions entirely to external non-VR platforms will unavoidably fragment users' immersive experience, and restricting such decisions to pre- or post-VR sessions disconnects users from the situational contexts necessary for their evaluation of privacy choices~\cite{nissenbaum2004privacy}. We argue that these considerations point to the need for multimodal privacy management solutions that leverage VR's unique properties to support users' privacy decision-making within immersive environments. 

\textit{Implications for Future Research:} Literature has proposed various VR-specific approaches for communicating privacy information and allowing users to convey privacy decisions (e.g.,~\cite{hadan2024privacy,rzayev2019notification,rakkolainen2021technologies}). Examples include delivering privacy notices through haptic feedback or in-situ displays on surrounding virtual objects, and gesture- and spatial-based controls that leverage the 3D interaction capabilities of VR technology~\cite{hadan2024privacy,rzayev2019notification,hadan2024cscw}. Scholars have further suggested that privacy design dimensions should be extended beyond visual, auditory, and haptic modalities to incorporate motion, neurofeedback, and even olfaction and gustation~\cite{hadan2024privacy,rakkolainen2021technologies}. However, most of these approaches remain conceptual or limited to in-lab prototypes, with little evidence of their effectiveness or applicability to real-world commercial VR applications. This is evident in our findings, where the privacy mechanisms we examined appeared to rely primarily on text and menu-based interactions, similar to those seen in computer and mobile systems (see~\autoref{subsec:findings-RQ3}). 

We encourage future VR researchers and industry practitioners to focus on expanding the design space for privacy in immersive systems~\cite{hadan2024privacy} and to examine how immersive embodiment, physical fatigue resulting from ergonomic issues, and trust cues may affect the feasibility and effectiveness of proposed mechanisms for supporting users’ privacy decisions. Operationalizing these approaches in real-world commercial VR contexts will likely require rethinking the physical and cognitive costs of engaging with VR hardware and privacy controls during immersion.

\subsection{Limitations}
We acknowledge several limitations of our study that point to opportunities for future research. 

First, our participants were exclusively experienced VR users \revision{from North America,} whose sustained engagement may reflect greater tolerance for manipulations but may potentially limit our findings. Future studies may examine users who discontinued VR use to reveal additional issues from privacy deceptive designs, \revision{and recruit participants across diverse geographic and cultural contexts to enhance the generalizability of these findings.}  

Second, our study relies on participants' self-reported perceptions and anticipated behaviors, which may differ from their actual behaviors in VR. Although we provided first-person video demonstrations to increase ecological validity, some scenarios may still depict situations that participants have not encountered or noticed in their own VR use, which can introduce recall or interpretation bias and affect the precision of their responses. \revision{In addition, our scenario materials were captured on a Meta Quest 3, and experiences may differ across devices. We also focused on single-user scenarios to avoid confounds from multi-user settings.} Future behavioral observation studies could complement our findings, though such work will require substantial care to protect participants against unintended privacy or manipulation risks.

\revision{Third, we situated participants in familiar VR applications to elicit their reflective interpretations while minimizing recall inaccuracies and to ethically avoid exposing them to real-world VR privacy violation. However, this approach could not capture \textit{in-situ} decision-making. We verified participants' prior VR application experience through survey questions to ensure ecological relevance, though our findings may not fully reflect their live VR interactions.} 

Fourth, we focused primarily on initial VR interactions, partly because current VR applications often avoided just-in-time privacy mechanisms to preserve immersion~\cite{hadan2024cscw}. Future studies may explore non-intrusive, just-in-time VR mechanisms that enable more accurate in-situ privacy decisions~\cite{hadan2024cscw,feng2021designspace}. 

Fifth, our findings suggest a potential relationship between the physical discomfort of VR headsets and users' responses to deceptive patterns (see~\autoref{subsec:findings-RQ3}). We argue that, similar to how VR's unbounded display and multimodality features could be leveraged as manipulative tactics~\cite{ramirez2021what,hadan2024deceived}, physical discomfort due to ergonomic issues may also be directly exploited as manipulative tactics as VR hardware matures. 

Lastly, our selected deceptive patterns are not exhaustive, and their frequent co-occurrence in practice~\cite{geronimo2020ui}  limited our ability to isolate perceptions of individual patterns. Our analysis focused on effects participants explicitly reported, potentially missing how VR-specific properties influence responses to unnoticed deception. We hope this study provides valuable insights from a user-centered perspective to encourage future work on more complex scenarios, such as multi-user interactions (e.g.,~\cite{ramirez2021what}) or additional VR mechanisms, to better understand how more diverse contexts impact users' concerns, interpretations, and responses. 

\section{Conclusion}
\label{sec:conclusion}
Our paper presents a scenario-based survey of 481 VR end-users' experiences and responses to privacy deceptive designs in commercial VR applications. We found that VR deceptive design can exploit cognitive vulnerabilities and bodily strain, \revision{a phenomenon we define as \textit{Ergonomic Susceptibility}.} Ergonomic issues from wearing VR headsets reduced users' patience and led to their hasty and careless decisions, especially when encountering patterns that involve cumbersome interactions. VR's realistic, sensory-rich experiences further contributed to users' susceptibility to manipulation by making invasive data requests appear immersion-preserving. While users implicitly recognized manipulation, their prior exposure to deceptive design in non-VR contexts led to privacy resignation. These findings emphasize the need to address ergonomics as a critical factor in future VR privacy design, and urge researchers, designers, and policymakers to develop ethical privacy management solutions that account for VR's unique multimodal, immersive, and ergonomic properties, building immersive virtual experiences that respect user privacy and mitigate manipulative data practices.

\begin{acks}
This project has been funded by the Office of the Privacy Commissioner of Canada (OPC); the views expressed herein are those of the author(s) and do not necessarily reflect those of the OPC, the University of Waterloo, nor the UWaterloo Games Institute. 

L. Zhang-Kennedy also acknowledge support from the Natural Sciences and Engineering Research Council of Canada (NSERC) Discovery Grant (\#RGPIN-2022-03353) and L. Nacke also acknowledge support from the Social Sciences and Humanities Research Council (SSHRC) INSIGHT Grant (\#435-2022-0476), Natural Sciences and Engineering Research Council of Canada (NSERC) Discovery Grant (\#RGPIN-2023-03705), and Canada Foundation for Innovation (CFI) John R. Evans Leaders Fund (\#41844). 

Thank you to the ACs for their effort in organizing the peer-review process, and thank you to the reviewers for their insightful feedback that helped us to improve the quality of this manuscript. We also thank post-doctoral researcher Dr. Reza Hadi Mogavi and Dr. Geneva Smith for their valuable feedback on the manuscript and statistical analyses prior to our submission. Screenshots in this manuscript were from the selected games and applications and fall under fair use.
\end{acks}

\bibliographystyle{ACM-Reference-Format}
\bibliography{02-References}

\appendix
\onecolumn 




\newpage
\clearpage
\section{Scenarios}
\label{app-sec:scenarios}

\renewcommand{\arraystretch}{1.5}  
\begin{table}[!ht]
\caption{Scenario description, corresponding VR application, design mechanism, and Incorporated privacy deceptive design patterns.}
\label{tab:LBWscenarios}
\centering
\resizebox{\columnwidth}{!}{%
\begin{tabular}{@{}p{0.2\textwidth}p{0.9\textwidth}@{}}
\toprule
\textbf{ID} & S1 \\ \midrule
\textbf{Sample Screenshot:} & \includegraphics[width=0.87\textwidth]{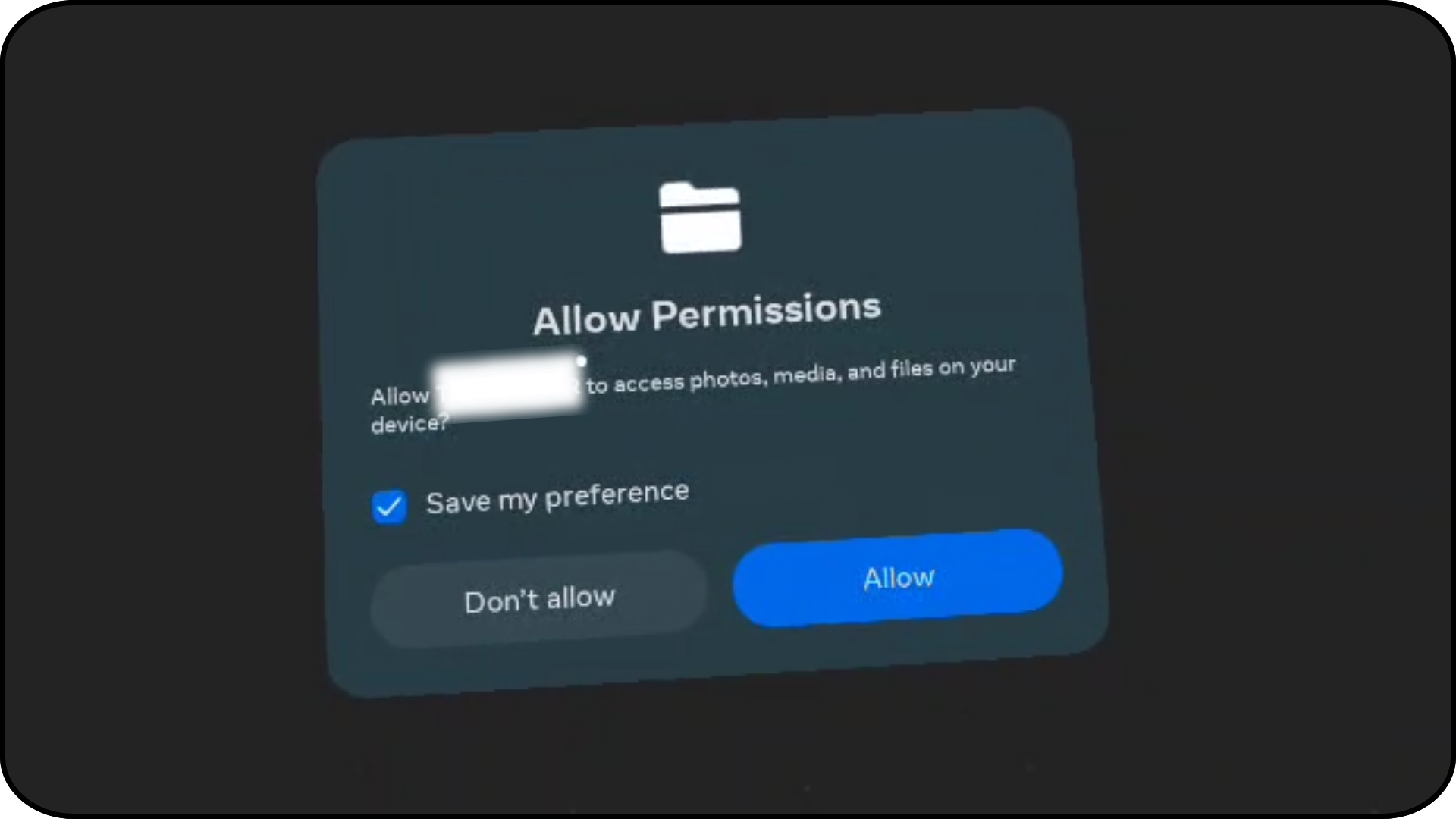} \\ \cdashlinelr{1-2}
\textbf{Scenario Description:} & Imagine you are about to use \textbf{VRChat}, a social VR application. Upon entering the application, you are immediately presented with a \textbf{Permission Request} asking whether to grant permissions for accessing your photos, media, and files. You must decide whether to approve or deny these permissions before proceeding further into the application. \\ \cdashlinelr{1-2}
\textbf{Deceptive Patterns:} & \begin{itemize}[leftmargin=1em,topsep=0em,]
    \item False Hierarchy/Visual Prominence: Highlight the ``Allow'' option that is invasive to user privacy.
    \item Bad Default: Set default options that benefit the company but risky for users' privacy.
    \item Privacy Zuckering: Trick users into thinking this permission is essential for the service.
    \item Hidden Information: Obscure crucial details of the purpose of data collection.
\end{itemize} \\ \cdashlinelr{1-2}
\textbf{Data Types Involved:} & Users' photos, media, and files on the VR device. \\ \cdashlinelr{1-2}
\textbf{Privacy Decision:} & Whether to allow the app to access photos, media, and files from the device, and whether to save permission preference. \\ \bottomrule
\end{tabular}%
}
\end{table}

\begin{table}[!ht]
\caption*{Table 4 Continued. Scenario description, corresponding VR application, design mechanism, and incorporated privacy deceptive design patterns.}
\label{tab:LBWscenarios-part2}
\centering
\resizebox{\columnwidth}{!}{%
\begin{tabular}{@{}p{0.2\textwidth}p{0.9\textwidth}@{}}
\toprule
\textbf{ID} & S2 \\ \midrule
\textbf{Sample Screenshot:} & \includegraphics[width=0.87\textwidth]{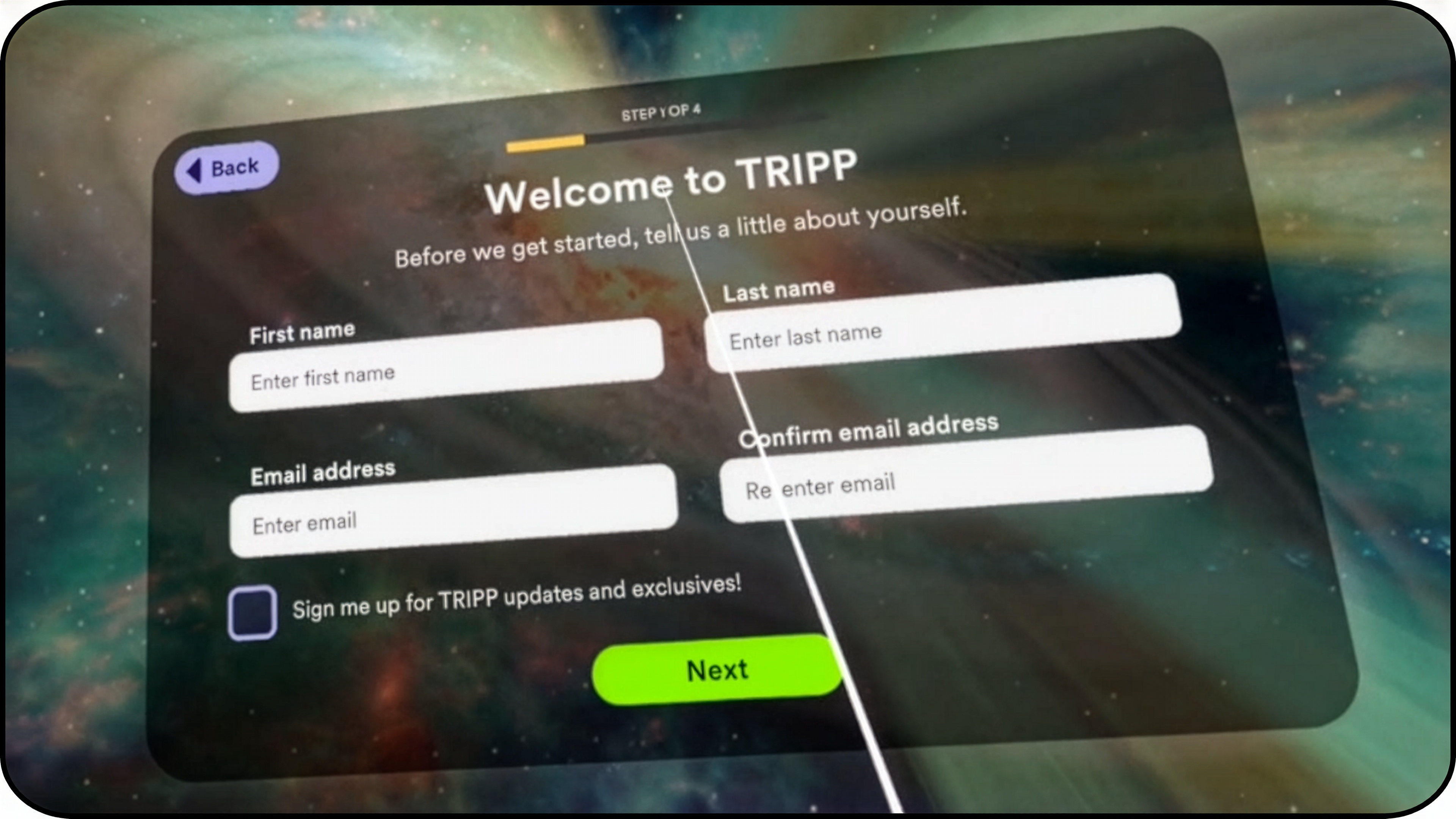} \\ \cdashlinelr{1-2}
\textbf{Scenario Description:} & Imagine you are about to use \textbf{TRIPP}, a VR wellness application. As you begin, you are presented with a \textbf{Registration Form}. Before accessing any functions, you are required to either enter your email address or register an account by providing your firstname and lastname to unlock all features. \\ \cdashlinelr{1-2}
\textbf{Deceptive Patterns:} & \begin{itemize}[leftmargin=1em,topsep=0em,]
    \item Forced Registration: Require account creation for tasks that should not need it, potentially extracting unnecessary personal data
    \item Privacy Zuckering: Trick users into thinking this permission is essential for the service.
    \item Positive or Negative Framing: Use vibrant green to communicate emotional safety on options that transmit user data.
\end{itemize} \\ \cdashlinelr{1-2}
\textbf{Data Types Involved:} & Users' full name and contact information. \\ \cdashlinelr{1-2}
\textbf{Privacy Decision:} & whether to provide personal identifiable information (e.g., full name) and contact information in the VR application. \\
\bottomrule
\end{tabular}%
}
\end{table}
\begin{table}[!ht]
\caption*{Table 4 Continued. Scenario description, corresponding VR application, design mechanism, and incorporated privacy deceptive design patterns.}
\label{tab:LBWscenarios-part3}
\centering
\resizebox{\columnwidth}{!}{%
\begin{tabular}{@{}p{0.2\textwidth}p{0.9\textwidth}@{}}
\toprule
\textbf{ID} & S3 \\ \midrule
\textbf{Sample Screenshot:} & \includegraphics[width=0.87\textwidth]{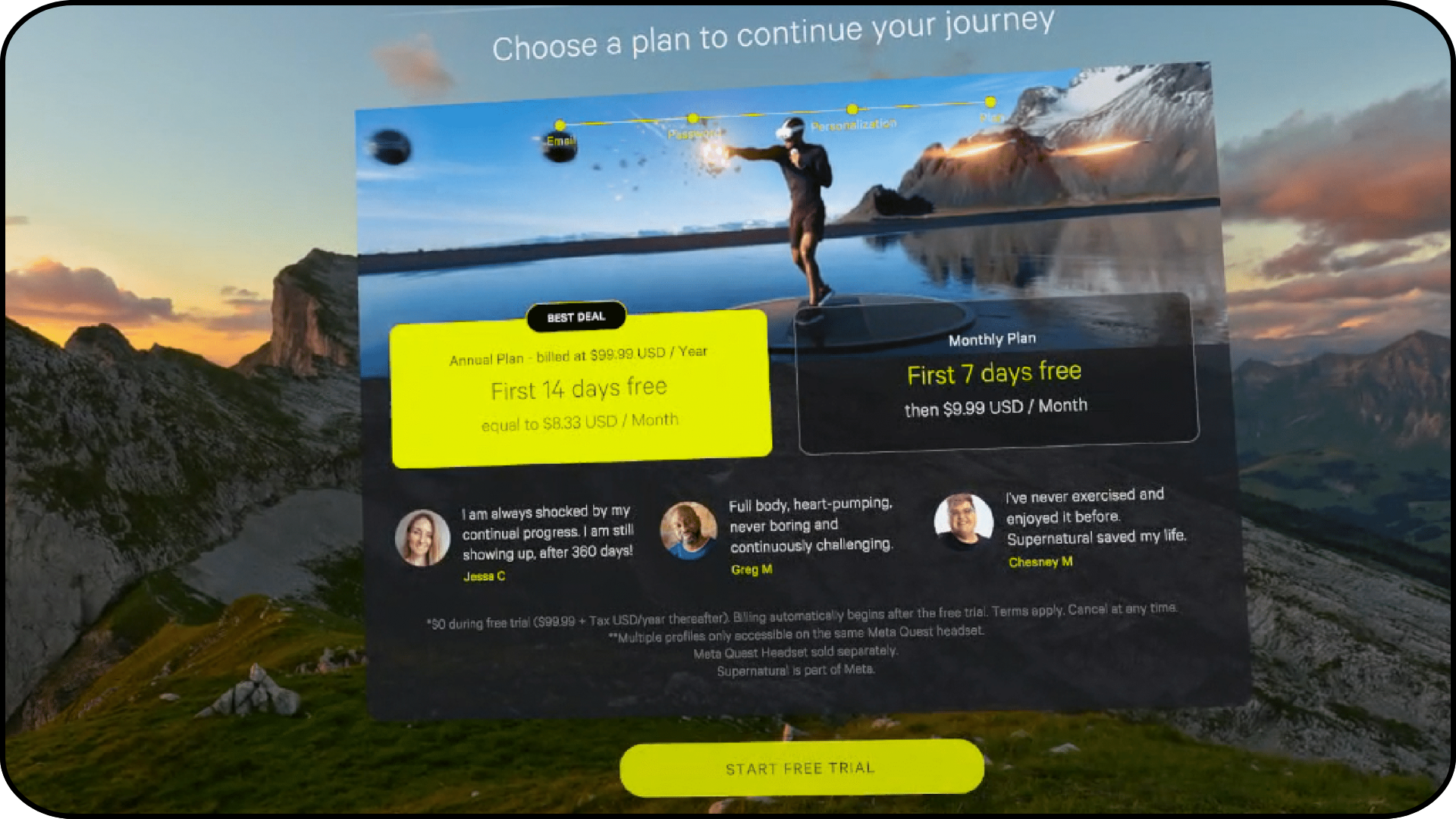} \\ \cdashlinelr{1-2}
\textbf{Scenario Description:} & Imagine you are about to use \textbf{Supernatural}, a VR fitness application that offers a subscription-based experience set in a photorealistic, immersive virtual landscape. As you start, you are presented with \textbf{Subscription Form} with two subscription plans and showcasing three testimonials. Before accessing any features, you are required to select and pay for a subscription. \\ \cdashlinelr{1-2}
\textbf{Deceptive Patterns:} & \begin{itemize}[leftmargin=1em,topsep=0em,]
    \item Endorsements and Testimonials: Use fabricated testimonials to influence users' purchase decisions.
    \item False Hierarchy/Visual Prominence: Highlight the acceptance option that transmits user data.
    \item Positive or Negative Framing: Use vibrant green to communicate emotional safety on options that transmit user data. 
\end{itemize} \\ \cdashlinelr{1-2}
\textbf{Data Types Involved:} & Users' payment information. \\ \cdashlinelr{1-2}
\textbf{Privacy Decision:} & Whether to disclose payment information to complete a subscription purchase. \\ \bottomrule
\end{tabular}%
}
\end{table}
\begin{table}[!ht]
\caption*{Table 4 Continued. Scenario description, corresponding VR application, design mechanism, and incorporated privacy deceptive design patterns.}
\label{tab:LBWscenarios-part4}
\centering
\resizebox{\columnwidth}{!}{%
\begin{tabular}{@{}p{0.2\textwidth}p{0.9\textwidth}@{}}
\toprule
\textbf{ID} & S4 \\ \midrule
\textbf{Sample Screenshot:} & \includegraphics[width=0.87\textwidth]{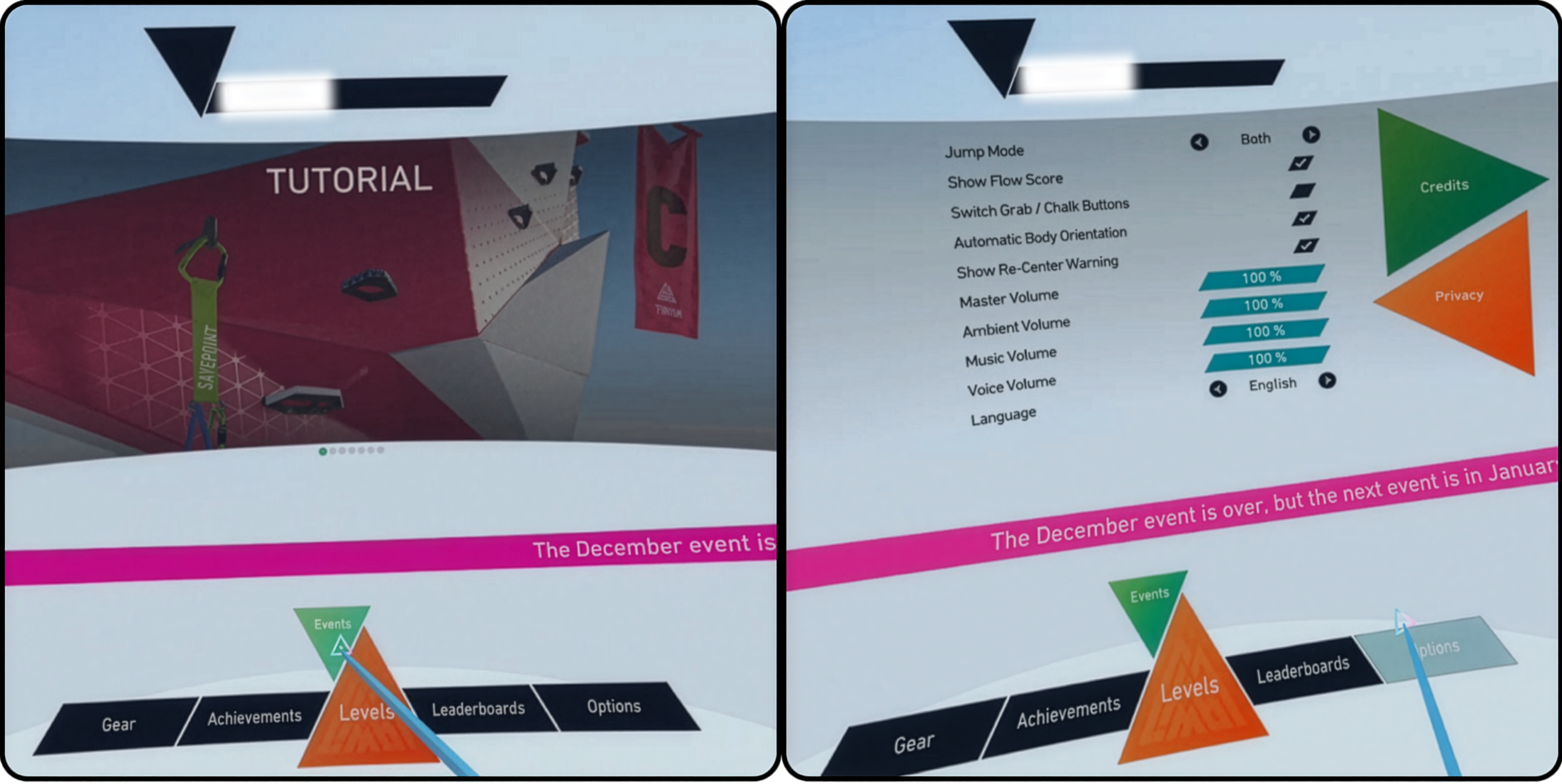} \\ \cdashlinelr{1-2}
\textbf{Scenario Description:} & Imagine you are using \textbf{The Climb 2}, a single-player VR game, and you are seeking to read its privacy policy. The main \textbf{Settings Menu} offers six distinct buttons, each leading to a different sub-menu. To look for the privacy policy, you must navigate to the ``options'' sub-menu, where you will see two buttons on the right and various application settings in the center. To access the privacy policy, you will need to select the ``privacy'' option on the right side. \\ \cdashlinelr{1-2}
\textbf{Deceptive Patterns:} & \begin{itemize}[leftmargin=1em,topsep=0em,]
    \item Privacy Maze: Bury privacy controls under layers of confusing menus. 
\end{itemize} \\ \cdashlinelr{1-2}
\textbf{Data Types Involved:} & N/A. \\ \cdashlinelr{1-2}
\textbf{Privacy Decision:} & Seek to access and review the privacy policy to understand how the application may store, share, or process personal or usage data. \\ \bottomrule
\end{tabular}%
}
\end{table}
\begin{table}[!ht]
\caption*{Table 4 Continued. Scenario description, corresponding VR application, design mechanism, and incorporated privacy deceptive design patterns.}
\label{tab:LBWscenarios-part5}
\centering
\resizebox{\columnwidth}{!}{%
\begin{tabular}{@{}p{0.2\textwidth}p{0.9\textwidth}@{}}
\toprule
\textbf{ID} & S5 \\ \midrule
\textbf{Sample Screenshot:} & \includegraphics[width=0.87\textwidth]{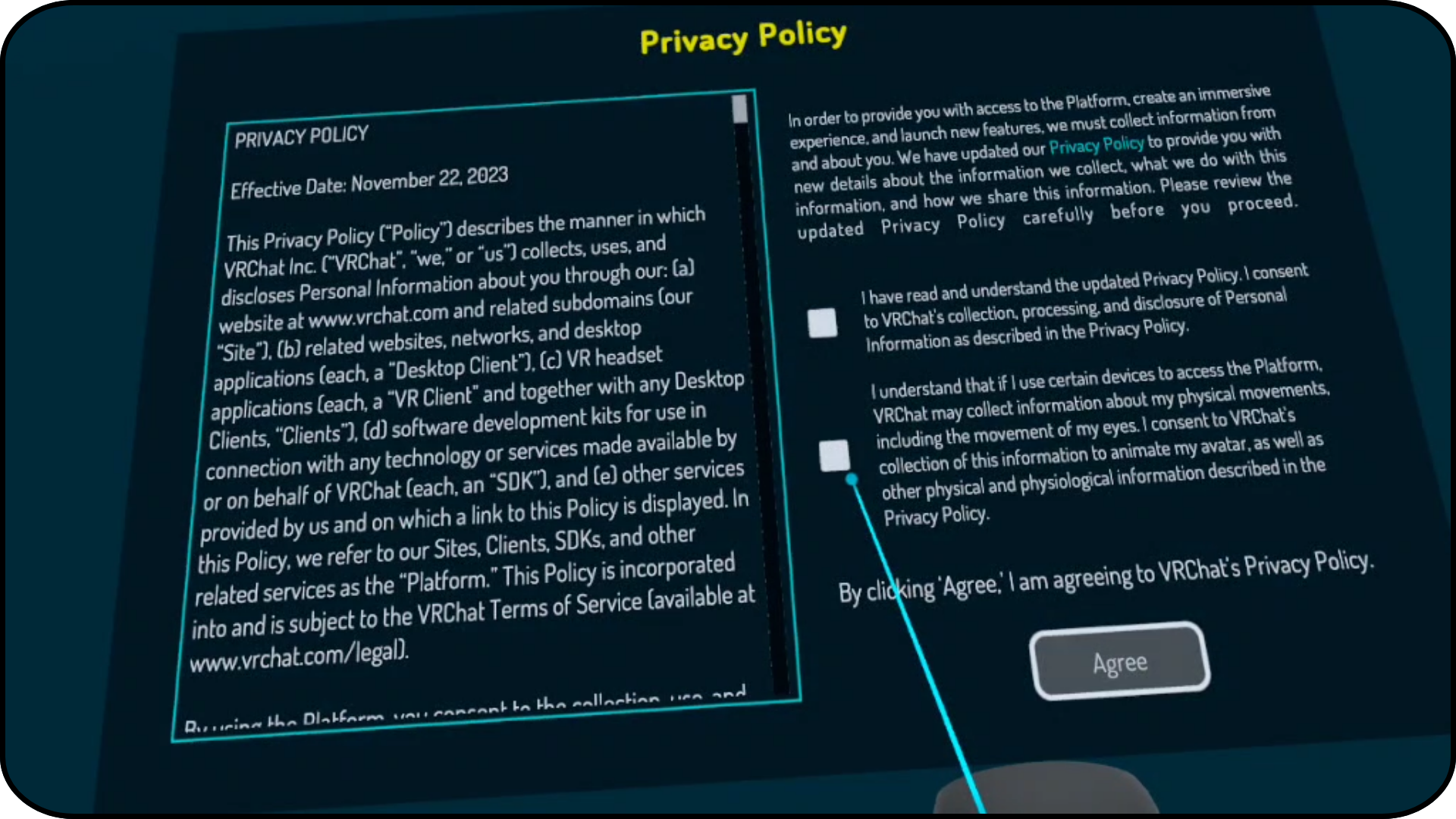} \\ \cdashlinelr{1-2}
\textbf{Scenario Description:} & Imagine you are about to use \textbf{VRChat}, a social VR application. This application requires you to review and consent to its privacy policy upon first entry. As you begin, you are presented with a \textbf{Consent Form} with a detailed privacy policy document that explains the application's data practices. You can scroll through the document to view the full content before checking the required boxes and clicking ``Agree.'' Consent must be given before you can access the main application. \\ \cdashlinelr{1-2}
\textbf{Deceptive Patterns:} & \begin{itemize}[leftmargin=1em,topsep=0em,]
    \item Complex \& Lengthy Language: Use hard-to-understand words and sentence structures to hinder users' understanding of privacy information.  
\end{itemize} \\ \cdashlinelr{1-2}
\textbf{Data Types Involved:} & N/A. \\ \cdashlinelr{1-2}
\textbf{Privacy Decision:} & Whether to agree to the terms that govern the collection, storage, and potential sharing of user information. \\ \bottomrule
\end{tabular}%
}
\end{table}
\begin{table}[!ht]
\caption*{Table 4 Continued. Scenario description, corresponding VR application, design mechanism, and incorporated privacy deceptive design patterns.}
\label{tab:LBWscenarios-part6}
\centering
\resizebox{\columnwidth}{!}{%
\begin{tabular}{@{}p{0.2\textwidth}p{0.9\textwidth}@{}}
\toprule
\textbf{ID} & S6 \\ \midrule
\textbf{Sample Screenshot:} & \includegraphics[width=0.87\textwidth]{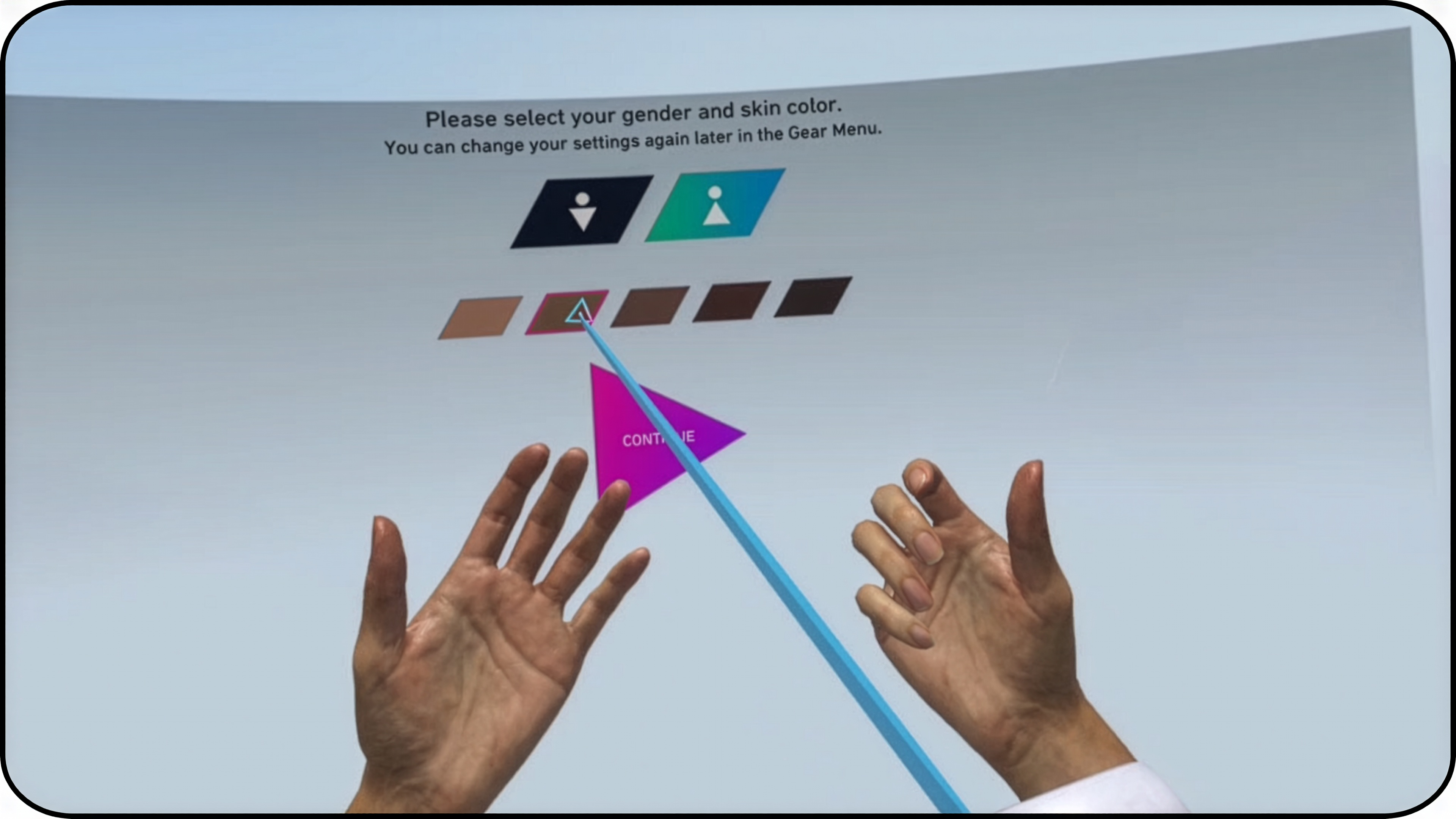} \\ \cdashlinelr{1-2}
\textbf{Scenario Description:} & Imagine you are about to use \textbf{The Climb 2}, a single-player VR game. As you start, you are prompted to customize your avatar. The \textbf{Avatar Creation Interface} asks you to select your gender and skin color, and you can see a photorealistic rendering of your avatar's hands from a first-person perspective. These choices will define how your avatar appears in the game world. \\ \cdashlinelr{1-2}
\textbf{Deceptive Patterns:} & \begin{itemize}[leftmargin=1em,topsep=0em,]
    \item Trick Questions: Use confusing wording to manipulate users' choices and lead them to provide real gender and skin color.  
\end{itemize} \\ \cdashlinelr{1-2}
\textbf{Data Types Involved:} & Users' gender and skin color. \\ \cdashlinelr{1-2}
\textbf{Privacy Decision:} & Provide personal information (e.g., gender, skin color) in the VR application. \\ \bottomrule
\end{tabular}%
}
\end{table}
\begin{table}[!ht]
\caption*{Table 4 Continued. Scenario description, corresponding VR application, design mechanism, and incorporated privacy deceptive design patterns.}
\label{tab:LBWscenarios-part7}
\centering
\resizebox{\columnwidth}{!}{%
\begin{tabular}{@{}p{0.2\textwidth}p{0.9\textwidth}@{}}
\toprule
\textbf{ID} & S7 \\ \midrule
\textbf{Sample Screenshot:} & \includegraphics[width=0.87\textwidth]{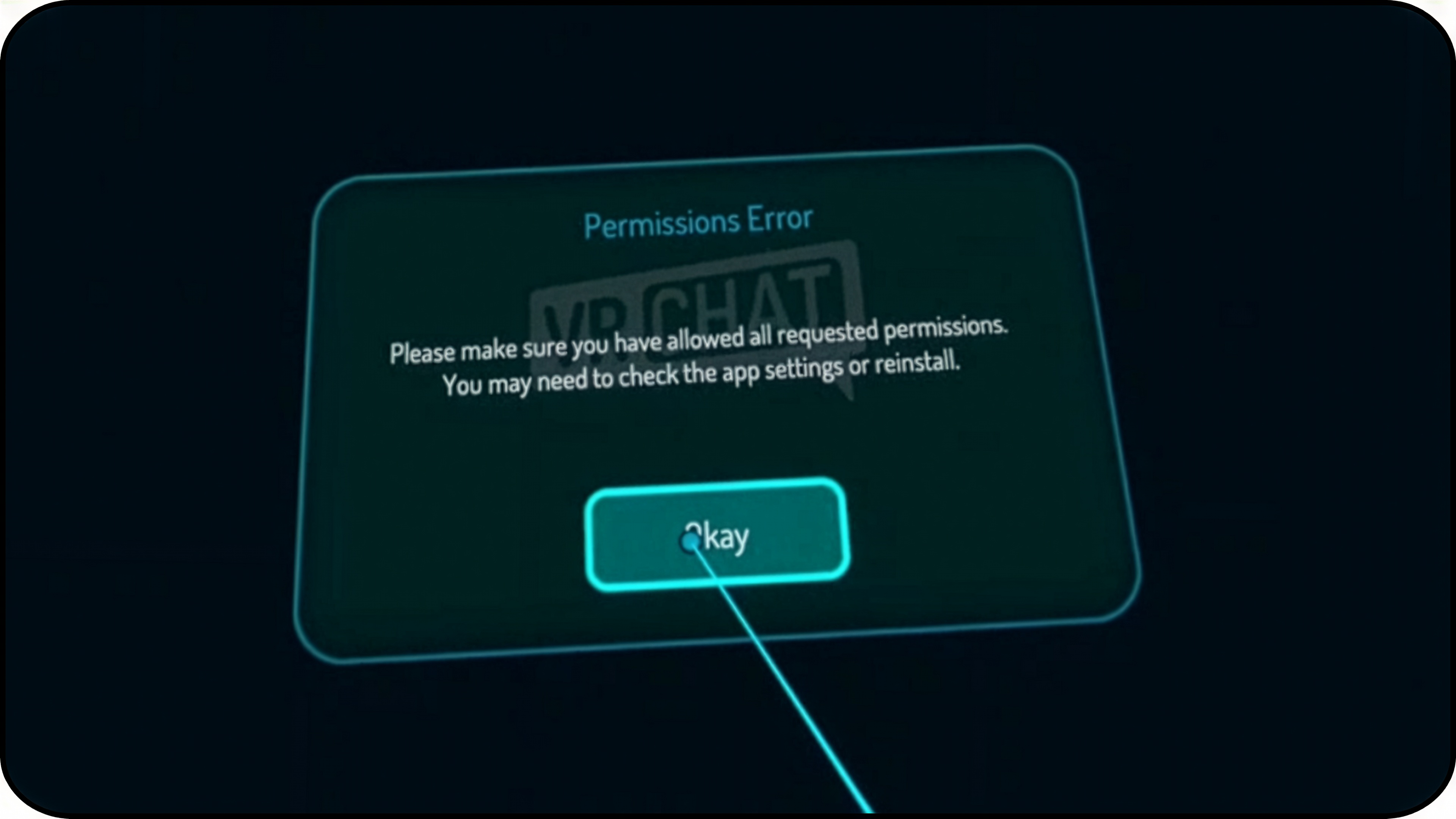} \\ \cdashlinelr{1-2}
\textbf{Scenario Description:} & Imagine you are about to use \textbf{VRChat}, a social VR application. Upon entry, you deny the application's request for access to your photos, media, and files. However, as you continue interacting with the application—completing the privacy policy consent, logging into your account, or trying to transfer to different virtual worlds—you are repeatedly interrupted by a ``permissions error'' \textbf{Notification} message. Each time, you must press ``okay'' to proceed. This Notification persists throughout your experience with VRChat. \\ \cdashlinelr{1-2}
\textbf{Deceptive Patterns:} & \begin{itemize}[leftmargin=1em,topsep=0em,]
    \item Nagging: Disrupt user workflow with repeatedly unwanted interruptions to push actions they would rather avoid.  
\end{itemize} \\ \cdashlinelr{1-2}
\textbf{Data Types Involved:} & N/A. \\ \cdashlinelr{1-2}
\textbf{Privacy Decision:} & Whether to restart the application to grant permission to avoid the nagging. \\ \bottomrule
\end{tabular}%
}
\end{table}
\begin{table}[!ht]
\caption*{Table 4 Continued. Scenario description, corresponding VR application, design mechanism, and incorporated privacy deceptive design patterns.}
\label{tab:LBWscenarios-part8}
\centering
\resizebox{\columnwidth}{!}{%
\begin{tabular}{@{}p{0.2\textwidth}p{0.9\textwidth}@{}}
\toprule
\textbf{ID} & S8 \\ \midrule
\textbf{Sample Screenshot:} & \includegraphics[width=0.87\textwidth]{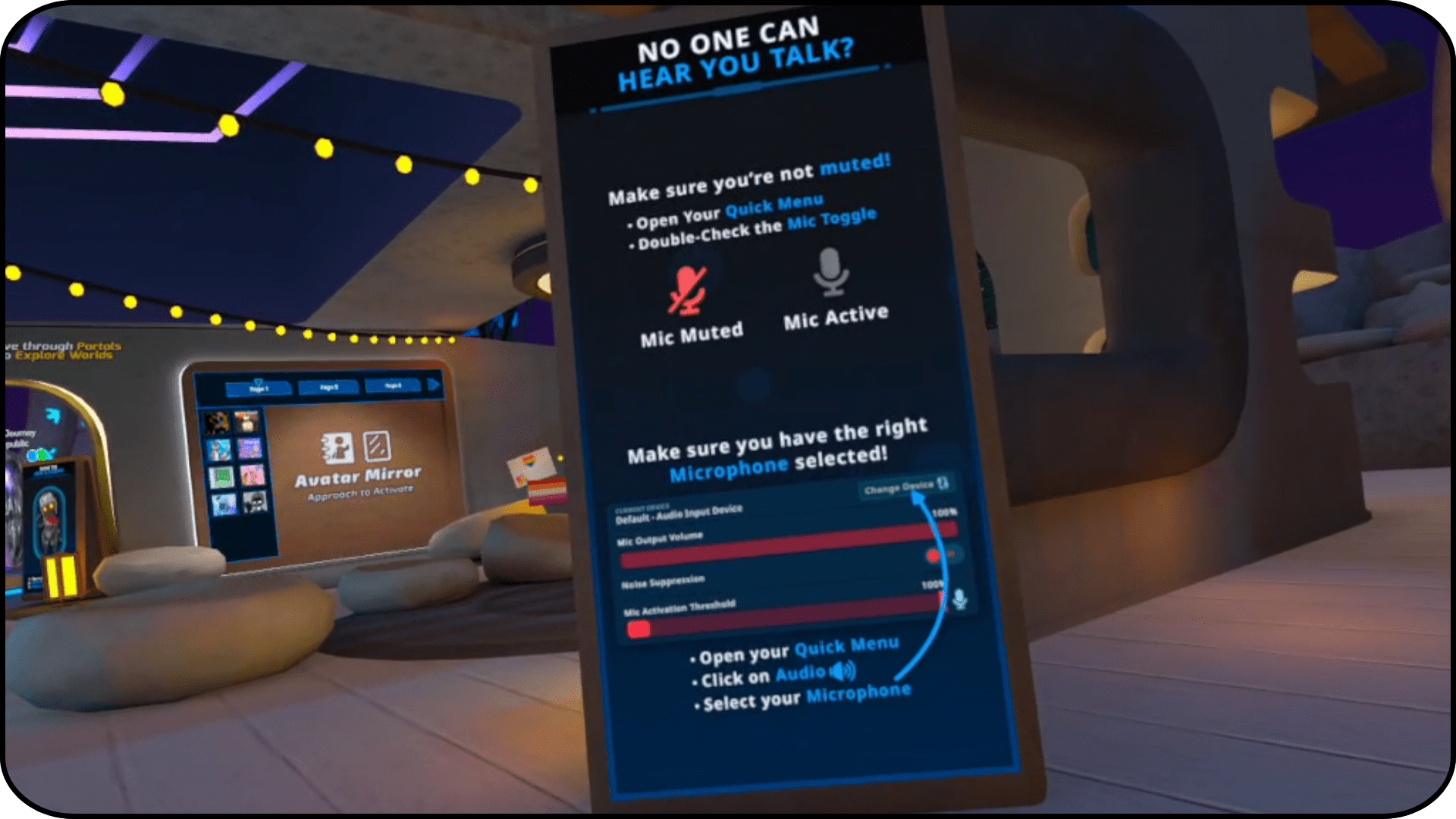} \\ \cdashlinelr{1-2}
\textbf{Scenario Description:} & Imagine you are about to use VRChat, a social VR application. Upon entering the first virtual world, you encounter a 3D virtual \textbf{Signboard} that gives you guidance about microphone settings. Following the guidance, you are able to open settings menu, locate and change your microphone settings, and mute and active your microphone. \\ \cdashlinelr{1-2}
\textbf{Deceptive Patterns:} & \begin{itemize}[leftmargin=1em,topsep=0em,]
    \item None.  
\end{itemize} \\ \cdashlinelr{1-2}
\textbf{Data Types Involved:} & Audio data. \\ \cdashlinelr{1-2}
\textbf{Privacy Decision:} & Whether to enable or disable their microphone, and allow the transmission of their voice to others in the VRChat. \\ \bottomrule
\end{tabular}%
}
\end{table}

\newpage
\clearpage
\section{Cumulative Link Mixed Model (CLMM) Regression Models}
\label{app-sec:statistics}

~\autoref{tab:CLMM} presents the full CLMM models obtained from using backward stepwise elimination. Positive estimates (or Odds Ratio>1) indicate a tendency towards greater awareness, perceived harm, or concern, whereas negative estimates (or Odds Ratio<1) indicate a tendency toward lower levels of these outcomes.

\begin{table*}[!th]
\vspace{10mm}
\caption{Multivariate Cumulative Link Mixed Model analyses of factors contributing to participants' ($N=481$) perceived benefits (1-strongly disagree to 5-strongly agree) of the design mechanisms for users like themselves (\varname{Benefit\_Me}), the application's developers and publishers (\varname{Benefit\_Dev}), and third-party stakeholders (e.g., marketing, legal services, analytics vendors) (\varname{Benefit\_Third-P}), and their perceived privacy impacts from interacting with the VR mechanisms (1-not at all, 5-significantly), with a random intercept per participant. We present the best-fitting models after backward stepwise selection.}
\centering
\resizebox{\textwidth}{!}{%
\begin{tabular}{@{}llcccrlllcccr@{}}
\multicolumn{5}{l}{DV=Benefit\_Me} & AIC=1542.97 &  & \multicolumn{5}{l}{DV=Benefit\_Dev} & AIC=1595.91 \\ \cline{1-6} \cline{8-13} 
\textbf{Predictor} & \textbf{Value} & \textbf{SE} & \textbf{z} & \textbf{\textit{P}} & \textbf{OR(95\%CI)} &  & \textbf{Predictor} & \textbf{Value} & \textbf{SE} & \textbf{z} & \textbf{\textit{P}} & \textbf{OR(95\%CI)} \\ \cline{1-6} \cline{8-13} 
\multicolumn{6}{l}{\textit{\textbf{Scenario\_Type}}} &  & \multicolumn{6}{l}{\textit{\textbf{Scenario\_Type}}} \\
Non-deceptive & \multicolumn{5}{l}{Reference} &  & Non-deceptive & \multicolumn{5}{l}{Reference} \\
Deceptive & -2.77 & 0.65 & -4.26 & \sig{$<.001$***} & 0.06 (0.02, 0.22) &  & Deceptive & 0.92 & 0.44 & 2.06 & \sig{$0.04$*} & 2.5 (1.05, 5.99) \\
\multicolumn{6}{l}{\textit{\textbf{VR\_ProExperience}}} &  & \multicolumn{6}{l}{\textit{\textbf{IUIPC\_Scores}}} \\
No & \multicolumn{5}{l}{Reference} &  & IUIPC\_Awareness & 0.76 & 0.18 & 4.31 & \sig{$<.001$***} & 2.14 (1.51, 3.02) \\
Yes & 1.54 & 0.52 & 2.97 & \sig{$<.001$***} & 4.64 (1.68, 12.8) &  & \multicolumn{6}{l}{\textit{\textbf{Deceptive Design Awareness \& Concern}}} \\
\multicolumn{6}{l}{\textit{\textbf{Star\_Rating}}} &  & Influence\_other & 0.22 & 0.16 & 1.35 & 0.18 & 1.25 (0.91, 1.71) \\
 & 1.21 & 0.28 & 4.33 & \sig{$<.001$***} & 3.35 (1.94, 5.79) &  & Concern\_other & -0.12 & 0.15 & -0.79 & 0.43 & 0.89 (0.66, 1.19) \\ \cline{8-13} 
\multicolumn{6}{l}{\textit{\textbf{IUIPC\_Scores}}} &  & \multicolumn{6}{l}{} \\
IUIPC\_Control & -0.39 & 0.19 & -2.02 & \sig{$0.04$*} & 0.68 (0.47, 0.99) &  & \multicolumn{5}{l}{DV=Privacy\_Impacts} & AIC=1278.48 \\ \cline{8-13} 
IUIPC\_Awareness & 1.00 & 0.27 & 3.72 & \sig{$<.001$***} & 2.71 (1.6, 4.57) &  & \textbf{Predictor} & \textbf{Value} & \textbf{SE} & \textbf{z} & \textbf{\textit{P}} & \textbf{OR(95\%CI)} \\ \cline{8-13} 
\multicolumn{6}{l}{\textit{\textbf{Deceptive Design Awareness \& Concern}}} &  & \multicolumn{6}{l}{\textit{\textbf{IUIPC\_Scores}}}  \\
Harm\_self & -0.06 & 0.19 & -0.30 & 0.76 & 0.95 (0.65, 1.37) &  & IUIPC\_Control & 1.22 & 0.64 & 1.91 & 0.06 & 3.4 (0.97, 11.94) \\
Concern\_other & -0.46 & 0.21 & -2.19 & \sig{$0.03$*} & 0.63 (0.42, 0.95) &  & \multicolumn{6}{l}{\textit{\textbf{Deceptive Design Awareness \& Concern}}} \\ \cline{1-6}
\multicolumn{6}{l}{} &  & Influence\_self & 1.88 & 0.58 & 3.23 & \sig{$<.001$***} & 6.53 (2.09, 20.39) \\ \cline{8-13} 
\multicolumn{5}{l}{DV=Benefit\_Third-P} & AIC=1673.91 &  & \multicolumn{6}{l}{} \\ \cline{1-6}
\textbf{Predictor} & \textbf{Value} & \textbf{SE} & \textbf{z} & \textbf{\textit{P}} & \textbf{OR(95\%CI)} &  & \multicolumn{6}{l}{\multirow{2}{*}{\begin{tabular}[c]{@{}l@{}} \small \textit{Note.} Significance are displayed as: \textcolor{red}{***}$P<.001$, \textcolor{red}{**}$P<.01$, \textcolor{red}{*}$P<.05$. DV=Dependent Variable. \\ \small SE=Standard Error. OR=Odds Ratio. CI=Confidence Interval. \end{tabular}} }\\ \cline{1-6}
\multicolumn{6}{l}{\textit{\textbf{Scenario\_Type}}} &  &  \multicolumn{6}{l}{ } \\
Non-deceptive & \multicolumn{5}{l}{Reference} &  & \multicolumn{6}{l}{ } \\
Deceptive & 2.73 & 0.58 & 4.69 & \sig{$<.001$***} & 15.35 (4.9, 48.14) &  & \multicolumn{6}{l}{ } \\ 
\multicolumn{6}{l}{\textit{\textbf{IUIPC\_Scores}}} &  & \multicolumn{6}{l}{\multirow{2}{*}{\begin{tabular}[c]{@{}l@{}} \\ \small  \end{tabular}}} \\
IUIPC\_Awareness & 0.76 & 0.22 & 3.51 & \sig{$<.001$***} & 2.14 (1.4, 3.27) &  & \multicolumn{6}{l}{}  \\ \cline{1-6}
\multicolumn{13}{l}{\begin{tabular}[c]{@{}l@{}} \small \textit{  }\\ \end{tabular}} \\ 
\end{tabular}%
}
\label{tab:CLMM}
\end{table*}

\newpage
\clearpage
\section*{Supplementary Materials}
\setcounter{section}{0}
\renewcommand{\thesection}{Supplementary Material \Alph{section}}
\renewcommand{\thesubsection}{\Alph{section}.\arabic{subsection}}

\begin{center}

\vspace*{1\baselineskip}

{\fontsize{14}{17}\selectfont\scshape Supplementary Materials}

\vspace{0.8\baselineskip}
\rule{80pt}{0.6pt}
\vspace{1.4\baselineskip}

{\fontsize{12}{15}\selectfont\itshape to accompany}

\vspace{0.7\baselineskip}

{\fontsize{14}{17}\selectfont\bfseries Rushed by Discomfort, Trapped by Immersion}

\vspace{1.1\baselineskip}

{\fontsize{12}{15}\selectfont\itshape Users' Experiences and Responses to Privacy\\Deceptive Design in Commercial VR Applications}

\vspace{1.6\baselineskip}

{\fontsize{12}{15}\selectfont Hadan, H., Valiquette, M., Nacke, L.\,E., \& Zhang-Kennedy, L. (2026)}

\vspace{0.5\baselineskip}

{\fontsize{11}{14}\selectfont\itshape In Proceedings of the 2026 ACM Conference on\\Designing Interactive Systems, pp.~1--34}

\vspace{0.6\baselineskip}

{\fontsize{11}{14}\selectfont\url{https://doi.org/10.1145/3800645.3812990}}

\vspace{1.4\baselineskip}
\rule{80pt}{0.6pt}
\vspace{0.7\baselineskip}

{\fontsize{14}{17}\selectfont\scshape Abstract}

\end{center}

\vspace{0.6\baselineskip}

\vspace{1.5\baselineskip}
\noindent\rule{\linewidth}{0.4pt}
\vspace{0.3\baselineskip}

\noindent\textit{Readers are encouraged to refer to our main paper for detailed descriptions of the study methodology and findings.}

\newpage
\clearpage
\section{Disclosure Statement for AI Use in Writing}

In accordance with ACM Publications Policy\footnote{Association for Computing Machinery. (July 6, 2023). ACM publications policy guidance for SIGCHI venues. Retrieved August 13, 2025, from ~\url{https://medium.com/sigchi/acm-publications-policy-guidance-for-sigchi-venues-87332173aad1}} and literature suggestion~\cite{hadan2024great}, we acknowledge our use of the Grammarly and Typingmind (GPT-4o and GPT-5 model) AI writing assistants. These tools were used to correct grammatical errors and, in a few cases, condense and improve sentence structure and suggest more suitable word choices using prompts: 
\begin{itemize}
    \item \textit{``fix any grammar errors in the following sentence and ensure a smooth flow [our human-written sentence].''}
    \item \textit{``give me 10 possible words/phrases that can more concisely and precisely reflect the idea in [our human-written phrase/sentence].''}
\end{itemize}

Our decision to use AI was based on our hope to achieve a concise and succinct writing while preserving the ``human-touch'' in our research writing~\cite{hadan2024great}. We did not use AI for data collection, analysis, or image generation. Qualitative analysis was manually conducted by our researchers using non-AI-supported version of Dovetail, and statistical analyses were conducted using R. Screenshots in this manuscript were captured using the Meta Quest 3 VR headset's screenshot function. Figures were created using Python seaborn package and were formatted using pre-built templates on Canva. Our research team manually reviewed, verified, and copy-edited the full manuscript. We take full responsibility for the content of the publication. 

\vspace{\baselineskip}   
\vspace{\baselineskip}   
\section{Data Availability Statement}

The data that support the findings of this study, including the scenario descriptions, screenshots, and short video clips are openly available in the Open Science Framework (OSF) at~\url{https://osf.io/axzve}.


\newpage
\clearpage
\section{Survey Questionnaire}

The following questionnaire was administered to all 481 participants during the study as described in the main paper.

\subsection{General Experience with VR Technology and Selected Applications}

\begin{enumerate}[label= \textbf{Q\arabic*:}]
    \item How long have you been using the VR equipment or applications?
    \begin{itemize}
        \item Less than 1 month
        \item 1 month to 6 months
        \item 6 months to 1 year
        \item 1 to 2 years
        \item More than 2 years
    \end{itemize}
    \item What is the main reason that encouraged you to use VR devices and applications? \textit{(select all that apply)}
    \begin{itemize}
        \item Entertainments (e.g., gaming, socializing)
        \item Healthcare (e.g., mental health diagnoses and treatment)
        \item Education or training (e.g., students, sports, military, medical procedures)
        \item Workplace requirement (e.g., remote work, staff connection, workplace functionality)
        \item Arts (e.g., virtual gallery, design \& prototyping, virtual graffiti)
        \item Other, please specify:
    \end{itemize}
    \item Have you ever worked in a professional role related to VR design, research, or development? \textit{(select all that apply)}
    \begin{itemize}
        \item VR Developer/Engineer
        \item VR Researcher
        \item VR User Experience (UX) or User Interface (UI) Designer
        \item VR Game Designer
        \item VR Product Manager
        \item VR Data Analyst
        \item Other, please specify:
        \item I have never worked in VR-related professional role.
    \end{itemize}
    \item You indicated in the screening questionnaire that you have experience using [the VR application]. How long did/have you been using this VR application?
    \begin{itemize}
        \item Less than 1 month
        \item 1 month to 6 months
        \item 6 months to 1 year
        \item 1 to 2 years
        \item More than 2 years 
    \end{itemize}
    \item In general, how often do/did you use this VR application?
    \begin{itemize}
        \item Everyday
        \item More than once a week
        \item Once a week
        \item More than once a month
        \item Once a month
        \item More than once a year
        \item Once a year
    \end{itemize}
    \item Overall, how would you rate your experience with this application? (5 stars = excellent experience)
    \item Please elaborate on specific aspects or features of this VR application that contribute to your star rating. (open-ended response)
    \item Please list the brand and model of the VR device you used to play this VR application. (open-ended response)
\end{enumerate}

\subsection{Perception and Experience with Privacy Deceptive Design Patterns}

Instruction: \textit{Next, we will show you a short video clip that demonstrates a VR design feature you may have encountered in The Climb 2.}

An Example Scenario: \textit{Imagine you are about to use \textbf{The Climb 2}, a single-player VR game. As you start, you are prompted to customize your avatar. The \textbf{Avatar Creation Interface} asks you to select your gender and skin color, and you can see a photorealistic rendering of your avatar's hands from a first-person perspective. These choices will define how your avatar appears in the game world.}

\begin{figure}[!ht]
\centering
  \includegraphics[width=0.7\textwidth]{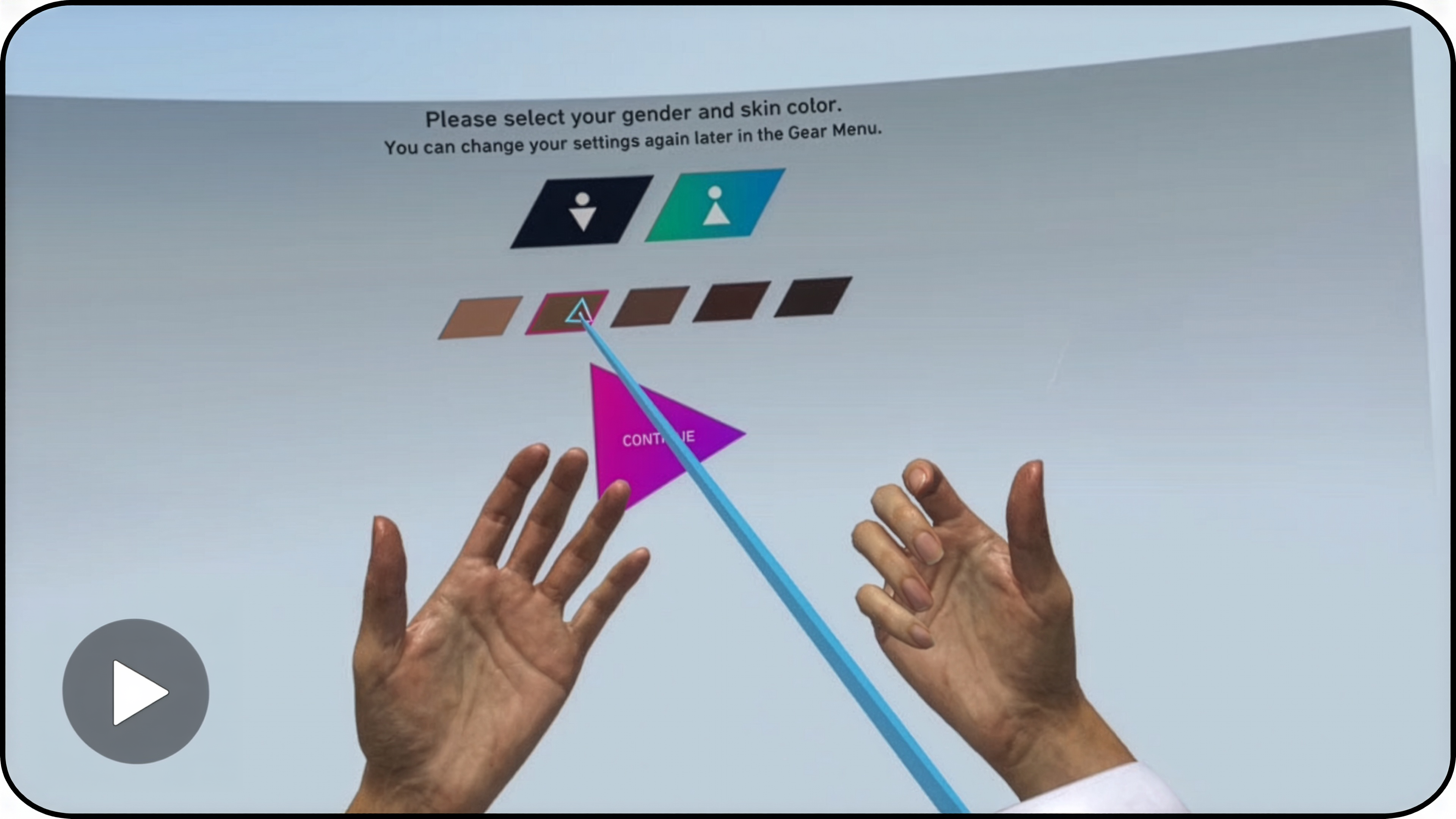}
  \caption{An example screenshot of the video clip.}
  \Description{Survey flow}
  \label{fig:sampleScreenshot}
\end{figure}

Instruction: \textit{Please answer the questions based on your personal experience with this avatar creation interface while using The Climb 2.}

\begin{enumerate}[resume,label= \textbf{Q\arabic*:}]
    \item Drawing from your previous experience using The Climb 2, what would be your thoughts and actions if you encountered this avatar creation interface? (open-ended response)
    \item Please briefly explain the reasons behind your thoughts and actions to this avatar creation interface. (open-ended response)
    \item When using this avatar creation interface, do you remember making any of the following decisions? \textit{(select all that apply)}
    \begin{itemize}
        \item I granted the application permission to access content (e.g., photos, media, files) stored on my VR device.
        \item I opted to remember my settings for future use of the application.
        \item I entered my personal details (e.g., name, email) into the application.
        \item I provided my payment information to subscribe to the application.
        \item I looked for and read through the application's data usage and protection policies.
        \item I do not remember making any decisions.
        \item I did not make any decisions.
        \item Other, please specify: [open space]
    \end{itemize}
    \item To what extent do you think your decision(s) might have impacted your personal privacy? (answered on 5-point scale(s) from 1-not at all to 5-significantly)
    \begin{itemize}
        \item Display the selected items in Q11. 
    \end{itemize}
    \item In which way do you think this decision might impact your privacy? (open-ended response)
    \begin{itemize}
        \item Display items in Q12 with a score $\geq$ 3. 
    \end{itemize}
\end{enumerate}

Instruction: \textit{The avatar creation interface you viewed includes \textbf{User Interface (UI) elements} that can facilitate your decision-making in your interaction.}

\textit{UI elements are the components of an interface that users interact with directly. Examples of UI elements include buttons, drop-down menus, sliders, text fields, checkboxes, toggles, and VR-specific features like realistic virtual objects or dynamic 3D environments.} 

\begin{figure}[!ht]
\centering
  \includegraphics[width=0.85\textwidth]{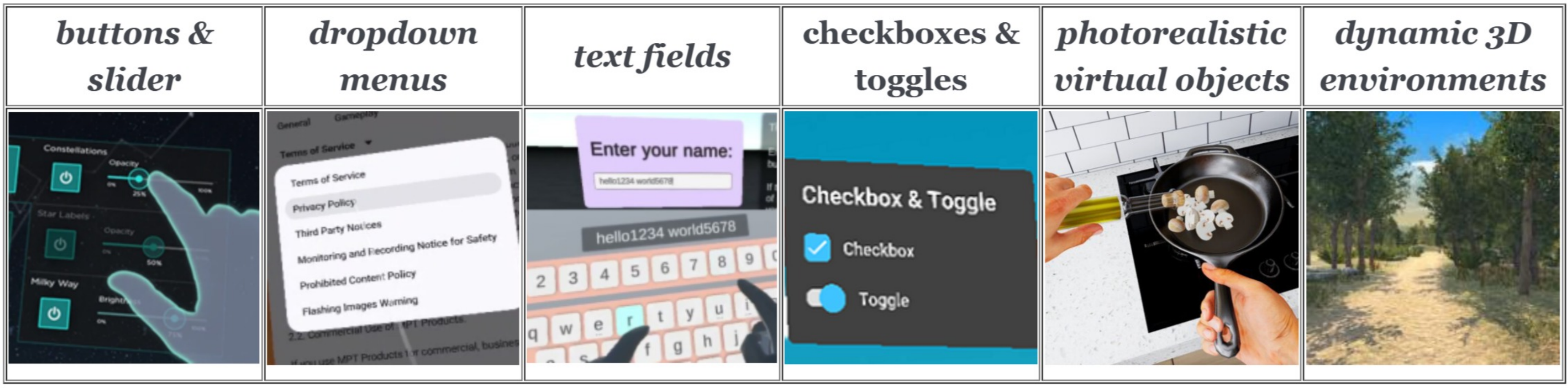}
  \caption{Example UI elements we used to support participants' understanding.}
  \Description{Survey flow}
  \label{fig:sampleIUelements}
\end{figure}

\begin{enumerate}[resume,label= \textbf{Q\arabic*:}] 
    \item Looking at the video again, did any specific design elements influence your decision-making process during your experience with this feature? If yes, please click and highlight the element(s). If design elements from other feature of this VR application influenced your decision-making with this avatar creation interface but are not shown in this screenshot, please select ``other'' and specify. (present a screenshot with design elements circled and labeled with corresponding names, options are customized for each scenario).
    \begin{itemize}
        \item Instruction Text
        \item Gender selection buttons
        \item Skin color selection buttons
        \item Photorealistic virtual hands
        \item Other, please specify:
    \end{itemize}
\end{enumerate}

The following question is repeated with each selected design element in Q14.

\begin{enumerate}[resume,label= \textbf{Q\arabic*:}]
    \item Please explain how this design element affected your decision during your own experience with this avatar creation interface. (open-ended response)
    \item Please indicate your level of agreements with the following statements (answered on 5-point Likert scales, from ``1-strongly disagree'' to ``5-strongly agree''): \textit{The decision or action I made as a result of interacting with this design mechanism is...}
    \begin{itemize}
        \item Beneficial to me.
        \item Beneficial to the application's developer/publisher.
        \item Beneficial to the third-party stakeholders (e.g., marketing, legal services, analytical vendors).
    \end{itemize}
\end{enumerate}

After repeating all the design elements, conclude this part of our study with Q17.

\begin{enumerate}[resume,label= \textbf{Q\arabic*:}]
    \item What is the intention of this [VR design mechanism] and its design elements? And why? (open-ended response)
\end{enumerate}

\subsection{General Concerns of Influence of Online Design and Privacy}
\label{app-subsec:general}

\begin{enumerate}[resume,label= \textbf{Q\arabic*:}]
    \item Please indicate your level of agreements with the following statements (answered on 5-point Likert scales, from -2-strongly disagree to 2-strongly agree) (adapted from~\cite{bongard2021definitely}): 
    \begin{itemize}
        \item The design of VR applications can influence [people’s/my] choices and behaviors on managing and sharing their personal information.
        \item VR applications that are designed to influence users can cause harm to [people's/my] personal information.
        \item I am worried about the influence of VR applications' design on [people's/my] choices and behaviors when it comes to managing and sharing their personal information. 
    \end{itemize}
    \item Internet Users’ Information Privacy Concerns (IUIPC) scale~\cite{IUIPCscale}. 
\end{enumerate}

\newpage
\clearpage
\section{Detailed Demographics}

\renewcommand{\arraystretch}{1}
\begin{table}[!ht]
\caption{Overview of Participants' demographics (N=481)}
\label{tab:detailed-demographics-part1}
\centering
\resizebox{\textwidth}{!}{%
\begin{tabular}{@{}llllllllll@{}}
\toprule
\textbf{Scenario(S)} & \textbf{S1 (n=62)} & \textbf{S2 (n=59)} & \textbf{S3 (n=61)} & \textbf{S4 (n=58)} & \textbf{S5 (n=64)} & \textbf{S6 (n=61)} & \textbf{S7 (n=59)} & \textbf{S8 (n=57)} & \textbf{All (n=481)}\\ \midrule
\multicolumn{10}{l}{{\textit{\textbf{Gender}}}} \\
Female & 27 (43.5\%) & 26 (44.1\%) & 33 (54.1\%) & 28 (48.3\%) & 26 (40.6\%) & 25 (41.0\%) & 27 (45.8\%) &  24 (42.1\%)  & 216 (44.9\%) \\
Male & 34 (54.8\%) & 32 (54.2\%) & 28 (45.9\%) & 29 (5$<$1\%) & 37 (57.8\%) & 36 (59.0\%) & 30 (50.8\%) & 32 (56.1\%) & 258 (53.6\%) \\
Non-binary / third gender & 1 (1.6\%) & 1 (1.7\%) & 0 (0\%) & 1 (1.7\%) & 0 (0\%) & 0 (0\%) & 2 (3.4\%) &  1 (1.8\%) & 6 (1.2\%) \\
Prefer not to say & 0 (0\%) & 0 (0\%) & 0 (0\%) & 0 (0\%) & 1 (1.6\%) & 0 (0\%) & 0 (0\%) &  0 (0\%) & 1 (0.2\%) \\
\multicolumn{10}{l}{}\\
\multicolumn{10}{l}{{\textit{\textbf{Age}}}} \\
Range (Min-Max) & 18-64 & 19-59 & 18-64 & 18-56 & 18-58 & 19-62 & 20-62 & 18-59 & 18-64 \\
Mean (SD) & 34.27 (11.15) & 34.61 (10.9) & 36.34 (11.02) & 31.33 (9.44) & 31.42 (9.81) & 33.28 (9.6) & 34.56 (10.63) & 31.02 (8.69) & 33.37 (10.29) \\
\multicolumn{10}{l}{}\\
\multicolumn{10}{l}{{\textit{\textbf{Education Level}}}} \\
High school or less & 13 (21.0\%) & 10 (16.9\%) & 8 (13.1\%) & 5 (8.6\%) & 13 (20.3\%) & 6 (9.8\%) & 5 (8.5\%) & 7 (12.3\%) & 67 (13.9\%) \\
Associate's degree & 6 (9.7\%) & 3 (5.1\%) & 11 (18.0\%) & 4 (6.9\%) & 7 (10.9\%) & 4 (6.6\%) & 5 (8.5\%) & 7 (12.3\%) & 47 (9.8\%) \\
Bachelor's degree & 35 (56.5\%) & 34 (57.6\%) & 33 (54.1\%) & 33 (56.9\%) & 33 (51.6\%) & 36 (59.0\%) & 38 (64.4\%) & 34 (59.7\%) & 276 (57.4\%) \\
Graduate's degree & 8 (12.9\%) & 12 (20.3\%) & 9 (14.8\%) & 16 (27.6\%) & 11 (17.2\%) & 15 (24.6\%) & 11 (18.6\%) & 91 (15.8\%) & 91 (18.9\%) \\
\multicolumn{10}{l}{}\\
\multicolumn{10}{l}{{\textit{\textbf{Country of Origin}}}} \\
Canada & 5 (8.1\%) & 29 (49.2\%) & 8 (13.1\%) & 7 (12.1\%) & 11 (17.2\%) & 4 (6.6\%) & 9 (15.3\%) &  22 (38.6\%)  & 95 (19.8\%) \\
United States of America & 57 (91.9\%) & 30 (50.8\%) & 53 (86.9\%) & 51 (87.9\%) & 53 (82.8\%) & 57 (93.4\%) & 50 (84.7\%) & 35 (61.4\%) & 386 (80.2\%) \\
\multicolumn{10}{l}{}\\
\multicolumn{10}{l}{{\textit{\textbf{Experience in Professional VR Roles}} (\varname{VR\_ProExperience})}} \\
VR Developer/Engineer & 1 (1.6\%) & 1 (1.7\%) & 5 (8.2\%) & 4 (6.9\%) & 1 (1.6\%) & 6 (9.8\%) & 4 (6.8\%) & 4 (7.0\%) & 26 (5.4\%) \\
VR Researcher & 3 (4.8\%) & 1 (1.7\%) & 8 (13.1\%) & 8 (13.8\%) & 10 (15.6\%) & 14 (23.0\%) & 7 (11.9\%) & 7 (12.3\%) & 58 (12.1\%) \\
VR Designer & 2 (3.2\%) & 3 (5.1\%) & 4 (6.6\%) & 8 (13.8\%) & 7 (10.9\%) & 8 (13.1\%) & 5 (8.5\%) & 3 (5.3\%) & 40 (8.3\%) \\
VR Product Manager & 4 (6.5\%) & 3 (5.1\%) & 8 (13.1\%) & 4 (6.9\%) & 3 (4.7\%) & 5 (8.2\%) & 8 (13.6\%) & 3 (5.3\%) & 38 (7.9\%) \\
VR Data Analyst & 3 (4.8\%) & 2 (3.4\%) & 6 (9.8\%) & 7 (12.1\%) & 6 (9.4\%) & 12 (19.7\%) & 5 (8.5\%) & 2 (3.5\%) & 42 (8.9\%) \\
Never & 53 (85.5\%) & 50 (84.7\%) & 47 (77.0\%) & 37 (63.8\%) & 47 (73.4\%) & 36 (59.0\%) & 45 (76.3\%) & 45 (78.9\%) & 360 (74.8\%) \\
Other & 0 (0.0\%) & 1 (1.7\%) & 1 (1.6\%) & 2 (3.4\%) & 0 (0.0\%) & 0 (0.0\%) & 1 (1.7\%) & 0 (0.0\%) & 5 (1.2\%) \\
\multicolumn{10}{l}{}\\
\multicolumn{10}{l}{{\textit{\textbf{IUIPC}}}} \\
\varname{IUIPC\_Control} &  &  & &  &  & & & & \\
Range (Min-Max) & 2-7 & 2-7 & 1-7 & 2-7 & 1-7 & 2-7 & 1-7 & 1-7 & 1-7 \\
Mean (SD) & 5.76 (0.96) & 5.53 (1.01) & 5.52 (1.09) & 5.57 (0.94) & 5.67 (1.02) & 5.63 (0.98) & 5.76 (0.9) & 5.96 (0.82) & 5.67 (0.96) \\
\varname{IUIPC\_Awareness} &  &  & &  &  & & & &  \\
Range (Min-Max) & 1-7 & 1-7 & 1-7 & 3-7 & 2-7 & 3-7 & 1-7 & 1-7 & 1-7 \\
Mean (SD) & 6.12 (1.02) & 6.2 (0.85) & 6.01 (1.06) & 6.07 (0.88) & 6.2 (0.91) & 6.07 (0.87) & 6.17 (0.82) & 6.36 (0.76) & 6.15 (0.90) \\
\varname{IUIPC\_Collection} &  &  & &  &  & & & &  \\
Range (Min-Max) & 1-7 & 1-7 & 1-7 & 1-7 & 1-7 & 1-7 & 1-7 & 1-7 & 1-7 \\
Mean (SD) & 5.38 (1.43) & 5.22 (1.61) & 5.26 (1.35) & 5.54 (1.08) & 5.21 (1.39) & 5.18 (1.33) & 5.61 (1.21) & 5.71 (1.13) & 5.39 (1.32) \\
\multicolumn{10}{l}{}\\
\multicolumn{10}{l}{{\textit{\textbf{Awareness and Concern about Deceptive Design}}}~\cite{bongard2021definitely}} \\
People &  &  & &  &  & & & & \\
Mean (SD) & 3.21 (1.13) & 2.77 (1.20) & 2.87 (1.28) & 2.94 (1.03) & 3.02 (1.20) & 2.79 (1.04) & 2.98 (1.32) & 3.18 (0.72) & 2.94 (1.17) \\
Median & 3 & 2.5 & 3 & 3 & 3 & 3 & 3 & 3 & 3 \\
Mode & 4 & 2 & 4 & 4 & 4 & 2 & 2 & 4 & 2 \\
Me &  &  & &  &  & & & & \\
Mean (SD) & 3.04 (1.18) & 2.66 (1.18) & 2.84 (1.13) & 3.05 (1.00) & 2.93 (1.25) & 2.67 (1.07) & 3.13 (1.29) & 2.89 (0.68) & 2.87 (1.15) \\
Median & 3 & 2 & 3 & 3 & 3 & 3 & 3 & 3 & 3 \\
Mode & 3 & 2 & 4 & 4 & 4 & 2 & 2 & 4 & 4 \\
 \bottomrule
\end{tabular}%
}
\end{table}
\begin{table}[!ht]
\caption*{Table 6 Continued. Overview of Participants' demographics (N=481)}
\label{tab:detailed-demographics-part2}
\centering
\resizebox{\textwidth}{!}{%
\begin{tabular}{@{}llllllllll@{}}
\toprule
\textbf{Scenario(S)} & \textbf{S1 (n=62)} & \textbf{S2 (n=59)} & \textbf{S3 (n=61)} & \textbf{S4 (n=58)} & \textbf{S5 (n=64)} & \textbf{S6 (n=61)} & \textbf{S7 (n=59)} & \textbf{S8 (n=57)} & \textbf{All (n=481)}\\ \midrule
\multicolumn{10}{l}{\textbf{\textit{Active Use of App}}} \\
No, I no longer actively use the app. & 14 (23\%) & 20 (34\%) & 14 (23\%) & 17 (29\%) & 13 (20\%) & 13 (21\%) & 8 (14\%) & 8 (14\%) & 107 (22\%) \\
Yes, I still actively use the app. & 48 (77\%) & 39 (66\%) & 47 (77\%) & 41 (71\%) & 51 (80\%) & 48 (79\%) & 51 (86\%) & 49 (86\%) & 374 (78\%) \\
\multicolumn{10}{l}{}\\
\multicolumn{10}{l}{\textbf{\textit{General Experience with VR Devices}} (\varname{VR\_Experience})} \\
Less than 6 months & 7 (11.3\%) & 4 (6.8\%) & 5 (8.2\%) & 11 (19.0\%) & 5 (7.8\%) & 2 (3.3\%) & 5 (8.5\%) & 3 (5.3\%) & 42 (8.7\%) \\
6 months to 1 year & 8 (12.9\%) & 8 (13.6\%) & 11 (18.0\%) & 5 (8.6\%) & 16 (25.0\%) & 12 (19.7\%) & 14 (23.7\%) & 5 (8.8\%) & 79 (16.4\%) \\
1 to 2 years & 17 (27.4\%) & 18 (30.5\%) & 20 (32.8\%) & 15 (25.9\%) & 17 (26.6\%) & 16 (26.2\%) & 15 (25.4\%) & 12 (21.1\%) & 130 (27.0\%) \\
2 to 5 years & 22 (35.5\%) & 24 (40.7\%) & 22 (36.1\%) & 22 (37.9\%) & 21 (32.8\%) & 20 (32.8\%) & 22 (37.3\%) & 23 (40.4\%) & 176 (36.6\%) \\
More than 5 years & 8 (12.9\%) & 5 (8.5\%) & 3 (4.9\%) & 5 (8.6\%) & 5 (7.8\%) & 11 (18.0\%) & 3 (5.1\%) & 14 (24.6\%) & 54 (11.2\%) \\
\multicolumn{10}{l}{}\\
\multicolumn{10}{l}{\textbf{\textit{Reason of VR Device Use}}} \\
Entertainments & 60 (96.8\%) & 56 (94.9\%) & 55 (90.2\%) & 51 (87.9\%) & 60 (93.8\%) & 55 (90.2\%) & 53 (89.8\%) & 55 (96.5\%) & 445 (92.5\%) \\
Healthcare & 7 (11.3\%) & 10 (16.9\%) & 18 (29.5\%) & 6 (10.3\%) & 3 (4.7\%) & 7 (11.5\%) & 7 (11.9\%) & 1 (1.8\%) & 58 (12.3\%) \\
Education or training & 16 (25.8\%) & 13 (22.0\%) & 18 (29.5\%) & 19 (32.8\%) & 15 (23.4\%) & 22 (36.1\%) & 11 (18.6\%) & 9 (15.8\%) & 123 (25.6\%) \\
Workplace requirement & 9 (14.5\%) & 4 (6.8\%) & 11 (18.0\%) & 10 (17.2\%) & 7 (10.9\%) & 13 (21.3\%) & 12 (20.3\%) & 5 (8.8\%) & 71 (14.8\%) \\
Arts & 21 (33.9\%) & 20 (33.9\%) & 13 (21.3\%) & 18 (31.0\%) & 19 (29.7\%) & 20 (32.8\%) & 20 (33.9\%) & 21 (36.8\%) & 152 (31.6\%) \\
Other & 2 (3.2\%) & 4 (6.8\%) & 6 (9.8\%) & 0 (0.0\%) & 2 (3.1\%) & 0 (0.0\%) & 3 (5.1\%) & 2 (3.5\%) & 19 (4.0\%) \\
\multicolumn{10}{l}{}\\
\multicolumn{10}{l}{\textbf{\textit{Experience with the VR Application}} (\varname{App\_Experience})} \\
Less than 1 month & 3 (4.8\%) & 9 (15.3\%) & 8 (13.1\%) & 5 (8.6\%) & 4 (6.3\%) & 6 (9.8\%) & 1 (1.7\%) & 0 (0.0\%) & 36 (7.5\%) \\
1 to 6 months & 12 (19.4\%) & 27 (45.8\%) & 19 (31.1\%) & 28 (48.3\%) & 16 (25.0\%) & 19 (31.1\%) & 13 (22.0\%) & 13 (22.8\%) & 147 (30.6\%) \\
6 months to 1 year & 12 (19.4\%) & 11 (18.6\%) & 15 (24.6\%) & 13 (22.4\%) & 13 (20.3\%) & 15 (24.6\%) & 18 (30.5\%) & 10 (17.5\%) & 107 (22.2\%) \\
1 to 2 years & 21 (33.9\%) & 9 (15.3\%) & 16 (26.2\%) & 9 (15.5\%) & 14 (21.9\%) & 15 (24.6\%) & 19 (32.2\%) & 15 (26.3\%) & 118 (24.5\%) \\
More than 2 years & 14 (22.6\%) & 3 (5.1\%) & 3 (4.9\%) & 3 (5.2\%) & 17 (26.6\%) & 6 (9.8\%) & 8 (13.6\%) & 19 (33.3\%) & 73 (15.2\%) \\
\multicolumn{10}{l}{}\\
\multicolumn{10}{l}{\textbf{\textit{Frequency of VR App Use}} (\varname{App\_Use\_Frequency})} \\
Everyday & 5 (8.1\%) & 4 (6.8\%) & 5 (8.2\%) & 2 (3.4\%) & 5 (7.8\%) & 4 (6.6\%) & 5 (8.5\%) & 5 (8.8\%) & 35 (7.3\%) \\
A few times a week & 23 (37.1\%) & 30 (50.8\%) & 35 (57.4\%) & 32 (55.2\%) & 37 (57.8\%) & 26 (42.6\%) & 30 (50.8\%) & 23 (40.4\%) & 236 (49.1\%) \\
Once a week & 10 (16.1\%) & 6 (10.2\%) & 5 (8.2\%) & 14 (24.1\%) & 7 (10.9\%) & 9 (14.8\%) & 5 (8.5\%) & 11 (19.3\%) & 67 (13.9\%) \\
A few times a month & 19 (30.6\%) & 14 (23.7\%) & 15 (24.6\%) & 10 (17.2\%) & 12 (18.8\%) & 19 (31.1\%) & 15 (25.4\%) & 14 (24.6\%) & 118 (24.5\%) \\
Once a month or less & 5 (8.1\%) & 5 (8.5\%) & 1 (1.6\%) & - & 3 (4.7\%) & 3 (4.9\%) & 4 (6.8\%) & 3 (5.3\%) & 24 (5.0\%) \\
no response & - & - & - & - & - & - & - & 1 (1.8\%) & 1 (0.2\%) \\
\multicolumn{10}{l}{}\\
\multicolumn{10}{l}{\textbf{\textit{Personal Rating of VR App Experience}} (\varname{Star\_Rating})} \\
1 Star & 0 (0.0\%) & 1 (1.7\%) & 1 (1.6\%) & 0 (0.0\%) & 0 (0.0\%) & 0 (0.0\%) & 1 (1.7\%) & 0 (0.0\%) & 3 (0.7\%) \\
2 Star & 3 (4.8\%) & 4 (6.8\%) & 2 (3.3\%) & 0 (0.0\%) & 1 (1.6\%) & 0 (0.0\%) & 1 (1.7\%) & 0 (0.0\%) & 11 (2.6\%) \\
3 Star & 11 (17.7\%) & 10 (16.9\%) & 11 (18.0\%) & 8 (13.8\%) & 10 (15.6\%) & 7 (11.5\%) & 9 (15.3\%) & 12 (21.1\%) & 78 (16.2\%) \\
4 Star & 31 (50.0\%) & 30 (50.8\%) & 27 (44.3\%) & 28 (48.3\%) & 38 (59.4\%) & 35 (57.4\%) & 36 (61.0\%) & 32 (56.1\%) & 257 (53.4\%) \\
5 Star & 16 (25.8\%) & 14 (23.7\%) & 20 (32.8\%) & 22 (37.9\%) & 15 (23.4\%) & 19 (31.1\%) & 12 (20.3\%) & 13 (22.8\%) & 131 (27.2\%) \\
no response & 1 (1.6\%) & - & - & - & - & - & - & - & 1 (0.2\%) \\  \bottomrule
\end{tabular}%
}
\end{table}

\newpage
\clearpage
\section{Descriptive Statistics}

\begin{table}[!h]
\caption{Descriptive statistics and Wilcoxon signed-rank test results comparing participants' ($N=481$) general (``others'') and personal (``self'') perspectives on VR design's influence on privacy decision-making, including awareness of influence, perceived harm, and level of concern~\cite{bongard2021definitely}, on a scale from ``1-strongly disagree'' to ``5-strongly agree''.}
\label{fig:ddawareness}
\begin{minipage}{0.80\textwidth}
\centering
\resizebox{\textwidth}{!}{%
\begin{tabular}{@{}llccclccp{0.17\textwidth}@{}}
\toprule
\textbf{Variable} & \textbf{Perspective} & \textbf{Median} & \textbf{Mode} & \textbf{Mean} & \textbf{SD} & \textbf{Min} & \textbf{Max} & \textbf{Wilcoxon test} \\ \midrule
VRD\_Influence  & General (\varname{VRD\_Influence\_other})  & 4 & 4 & 3.34 & 1.08 & 1 & 5 & W = 11081, \\
 & Personal (\varname{VRD\_Influence\_self})&  3 & 4 & 3.20 & 1.11 & 1 & 5 &  \sig{P = .001}, \\
& & & & & & & & r=-.15 \\  \cdashlinelr{1-9}
VRD\_Harm & General (\varname{VRD\_Harm\_other}) &  3 & 2  & 2.84 & 1.19 & 1 & 5 & W = 8849, \\
 & Personal (\varname{VRD\_Harm\_self}) &  3 & 2 & 2.82 & 1.16 & 1 & 5 & P = .931, \\ 
 & & & & & & & & r=-.004 \\ \cdashlinelr{1-9}
VRD\_Concern & General (\varname{VRD\_Concern\_other}) &  3 & 2  & 2.77 & 1.18 & 1 & 5 & W = 11528,  \\
 & Personal (\varname{VRD\_Harm\_other}) &  2 & 2  & 2.64 & 1.17 & 1 & 5 & \sig{P = .003}, \\ 
 & & & & & & & & r=-.13 \\ \bottomrule
\multicolumn{9}{l}{\small \textit{Note. *P adjusted with Bonferroni correction. SD=Standard Deviation, W=test statistic, r=effect size.}}
\end{tabular}%
}
\end{minipage}
\end{table}

\begin{table}[!h]
\caption{Descriptive statistics of participants' ($N=481$) perceptions of VR design mechanisms' benefits for users like themselves, for application developers or publishers, and for third party stakeholders (e.g., marketing, legal services, analytical vendors), on a scale from ``1-strongly disagree'' to ``5-strongly agree''.}
\label{tab:benefit-descriptive}
\centering
\resizebox{0.45\textwidth}{!}{%
\begin{tabular}{@{}llccc@{}}
\toprule
\textbf{Scenario} & \textbf{Variable} & \multicolumn{1}{l}{\textbf{Mean (SD)}} & \multicolumn{1}{l}{\textbf{Median}} & \multicolumn{1}{l}{\textbf{Mode}} \\ \midrule
S1 & \varname{Benefit\_me} & 3.65 (1.17) & 4 & 3 \\
 & \varname{Benefit\_dev} & 3.91 (0.99) & 4 & 4 \\
 & \varname{Benefit\_Third-P} & 3.79 (1.05) & 4 & 4 \\
 &  &  &  &  \\
S2 & \varname{Benefit\_me} & 3.96 (0.96) & 4 & 4 \\
 & \varname{Benefit\_dev} & 4.15 (0.91) & 4 & 5 \\
 & \varname{Benefit\_Third-P} & 3.86 (1.02) & 4 & 4 \\
 &  &  &  &  \\
S3 & \varname{Benefit\_me} & 4.3 (0.81) & 4 & 4 \\
 & \varname{Benefit\_dev} & 4.19 (0.66) & 4 & 4 \\
 & \varname{Benefit\_Third-P} & 4.1 (0.81) & 4 & 4 \\
 &  &  &  &  \\
S4 & \varname{Benefit\_me} & 4.06 (0.8) & 4 & 4 \\
 & \varname{Benefit\_dev} & 3.83 (0.73) & 4 & 4 \\
 & \varname{Benefit\_Third-P} & 3.65 (0.81) & 4 & 3 \\
 &  &  &  &  \\
S5 & \varname{Benefit\_me} & 3.86 (1.08) & 4 & 4 \\
 & \varname{Benefit\_dev} & 4.07 (0.6) & 4 & 4 \\
 & \varname{Benefit\_Third-P} & 4.04 (0.79) & 4 & 4 \\
 &  &  &  &  \\
S6 & \varname{Benefit\_me} & 4.21 (0.73) & 4 & 4 \\
 & \varname{Benefit\_dev} & 4.03 (0.66) & 4 & 4 \\
 & \varname{Benefit\_Third-P} & 3.87 (0.78) & 4 & 4 \\
 &  &  &  &  \\
S7 & \varname{Benefit\_me} & 2.67 (1.45) & 2 & 1 \\
 & \varname{Benefit\_dev} & 4.33 (0.82) & 5 & 5 \\
 & \varname{Benefit\_Third-P} & 4.07 (0.96) & 4 & 5 \\ 
  &  &  &  &  \\
S8 & \varname{Benefit\_me} & 4.34 (0.90) & 4 & 5 \\
 & \varname{Benefit\_dev} & 3.76 (0.93) & 4 & 4 \\
 & \varname{Benefit\_Third-P} & 3.18 (1.11) & 3 & 3 \\ 
 \bottomrule
\end{tabular}%
}
\end{table}

\begin{table}[!ht]
\caption{Descriptive statistics of participants' ($N=481$) perceived privacy impact of decisions they made during their interactions with the VR design mechanisms, ``1-not at all'' to ``5-significantly''.}
\label{tab:privacy-impact-descriptive}
\centering
\resizebox{0.7\textwidth}{!}{%
\begin{tabular}{@{}llccc@{}}
\toprule
\textbf{Scenario} & \textbf{Privacy Decision} & \multicolumn{1}{l}{\textbf{Mean (SD)}} & \multicolumn{1}{l}{\textbf{Median}} & \multicolumn{1}{l}{\textbf{Mode}} \\ \midrule
S1 & I granted the application permission & 2.68 (1.32) & 3 & 1 \\
S1 & I opted to save my preferences & 2.73 (1.34) & 2.5 & 2 \\
S2 & I entered my personal details & 2.85 (1.35) & 3 & 4 \\
S3 & I provided my payment information & 3.11 (1.37) & 3 & 4 \\
S4 & I searched and read through the privacy policy & 2.69 (1.28) & 3 & 2 \\
S5 & I granted the application permission & 3.1 (1.24) & 3 & 3 \\
S5 & I searched and read through the privacy policy & 3.36 (1.54) & 4 & 5 \\
S6 & I entered my personal details & 3.56 (1.28) & 4 & 4 \\
S7 & I granted the application permission & 3.48 (1.06) & 4 & 4 \\ 
S8 & I updated my microphone settings & 2.89 (1.43) & 3 & 1 \\ 
\bottomrule
\end{tabular}%
}
\end{table}

\newpage
\clearpage
\section{Thematic Analysis Codebooks}
\label{app-sec:codebook}

\renewcommand{\arraystretch}{1.2}
\begin{table}[!h]
\caption{This codebook presents the codes and themes we synthesized from participants' responses to the open-ended question ``Please elaborate on specific aspects or features of this VR application that contribute to your star rating'' (Q7).}
\label{tab:codebook-starRating}
\centering
\resizebox{\textwidth}{!}{%
\begin{tabular}{@{}llcccccp{0.43\textwidth}@{}}
\toprule
\multicolumn{2}{l}{\multirow{2}{*}{\textbf{\textit{Theme}/Code}}} & \multicolumn{5}{c}{\textbf{Number of participants}} & \multirow{2}{*}{\textbf{Example quotes from participants' responses}} \\ \cmidrule(lr){3-7}
\multicolumn{2}{l}{} & \textbf{VRChat} & \textbf{TRIPP} & \textbf{Supernatural} & \textbf{The Climb2} & \textbf{Total} &  \\ \midrule
\multicolumn{8}{l}{\textbf{\textit{Positive VR Experience}}} \\
 & Good experience \& game modes & 144 & 50 & 47 & 91 & 332 & ``I really liked being able to create my avatar that has a lot of my features.'' \\
 & Good video, audio, function & 55 & 0 & 0 & 0 & 55 & ``great graphics, liked voice choice, enjoyed the ability to mix and match the visual stimulus to different meditations.'' \\
 & Benefit me outside of the app & 3 & 10 & 1 & 1 & 15 & ``It has also completely changed, for the better, my sleep routine.'' \\
 & Nostalgia & 2 & 0 & 0 & 0 & 2 & ``It reminds me of the days of chatting with others on PalTalk chat rooms or IMVU.'' \\ \cdashlinelr{1-8}
\multicolumn{8}{l}{\textbf{\textit{Negative VR Experience}}} \\
 & Technical, accessibility issues & 37 & 12 & 9 & 12 & 70 & ``I find the experience to come off a big clunky and unrefined.'' \\
 & Toxicity, harassment, social annoyances & 43 & 0 & 0 & 0 & 43 & ``Other users were negative to me and made the experience not as enjoyable.'' \\
 & Misalignment with personal & 11 & 6 & 6 & 10 & 33 & ``I got bored after the initial period of novelty ran out.'' \\
 & Limited content & 2 & 7 & 4 & 9 & 22 & ``It simply becomes a bit boring and uneventful after prolonged play time.'' \\
 & Kids vs. adults conflicts & 9 & 0 & 0 & 0 & 9 & ``It also gets filled with children a lot.'' \\
 & Problematic subscription \& pricing & 0 & 2 & 7 & 0 & 9 & ``I guess I was expecting more out of the system. The monthly subscription is higher than I would have expected.'' \\
 & Expected longer trial period & 0 & 0 & 1 & 0 & 1 & ``I know there's a learning curve, but the free trial wasn't long and I didn't want to pay for it when I was being frustrated by it.'' \\
 & Free-app-justifies & 1 & 0 & 0 & 0 & 1 & ``It's a free app.'' \\ 
 & No response & 0 & 0 & 1 & 0 & 1 &  \\ \bottomrule
\end{tabular}%
}
\end{table}
\renewcommand{\arraystretch}{1}

\renewcommand{\arraystretch}{1.2}
\begin{table}[!t]
\caption{This codebook presents the codes and themes we synthesized from participants' responses to the open-ended question ``Drawing from your previous experience using [the VR application], what would be your thoughts and actions if you encountered this [VR design mechanism]?'' (Q9) and ``Please briefly explain the reasons...'' (Q10).}
\label{tab:codebook-Q19thoughtsreactions}
\centering
\resizebox{\textwidth}{!}{%
\begin{tabular}{@{}llcp{0.6\textwidth}@{}}
\toprule
\textbf{Scenario (S)} & \textbf{Themes} & \textbf{n (\%)} &\textbf{Example Quotes} \\ \midrule
S1 & Perceived Trust and Legitimacy & 34 (55\%) & \textit{``Often times you have to allow app access in order to have full usage.''} \textit{``I think they wanted to get permission before the game play start to avoid issues within the game.''}\\
S1 & Concerns \& Reluctance & 32 (52\%) & \textit{``Not sure why it needed access, but i gave it access to run the game properly.''}  \textit{``I was afraid that they wanted to steal all of my personal information.''}\\
S1 & Coping Strategies \& Desires & 13 (21\%) & \textit{``Just let me select the specific access I want to give.''} \textit{``I think they need to explain the uses of it and maybe have a limited option?''}\\
S1 & Design normalization \& Indifference & 19 (31\%) & \textit{``I don't have files on my device that I don't want this program to see.''} \textit{``We already allowed this with our phones. So I pretty much have enough trust with this.''}\\
S1 & Lack of Choice & 5 (8\%) & \textit{``I didn't want to be rejected from the platform.''} \textit{``So I don't have to do permission again.''} \\
& & \\
S2 & Perceived Trust and Legitimacy & 35 (39\%) & \textit{``Entered in my information as prompted on the screen. Seemed like a generic account creation.''} \textit{``I filled in my information as requested.''}\\
S2 & Concerns \& Reluctance & 23 (39\%) & \textit{``I was fairly concerned of data breaches and I did not feel there is much reason for them to have my full name.''} \textit{``I wondered what would happen to me information.''}\\
S2 & Coping Strategies \& Desires & 13 (22\%) & \textit{``I really dislike when an app asks for login information before even seeing what it's about. There has to be a better way to go about this.''} \textit{``I wondered why my name was required. I chose to enter an abbreviated name and used a secondary email address.''}\\
S2 & Design normalization \& Indifference & 27 (46\%) & \textit{``I have previously entered my info in a similar situation.''} \textit{``I already had an account for other applications, why not this too.''}\\
& & & \\
S3 & Perceived Trust and Legitimacy & 34 (56\%) & \textit{``I was hooked and ready to sign up!''} \textit{``I have heard so many good things about it and a lot of people love the program.''}\\
S3 & Concerns \& Reluctance & 16 (26\%) & \textit{``I was scared of forgetting to cancel and getting charged \$99.''} \textit{``they could be fake ads which those appeared to be because the photos alongside the ads looked too polished in my opinion.''}\\
S3 & Coping Strategies \& Desires & 40 (66\%) & \textit{``The 7 day trial was just so short. If it was 30 and 14 days, that would seem more attractive to me.''} \textit{``I compared it to other online fitness programs.''}\\
S3 & Design normalization \& Indifference & 3 (5\%) & \textit{``It's part of those things one encounter in this kind of things.''} \textit{``I did not care.''}\\
S3 & Lack of Choice & 1 (2\%) & \textit{``Heavily geared towards getting you sign up for longer plan.''}\\
& & & \\
S4 & Perceived Trust and Legitimacy & 23 (40\%) & \textit{``It was rather easy to find.''} \textit{``Privacy policy is thorough and easily understandable.''}\\
S4 & Concerns \& Reluctance & 5 (9\%) & \textit{``This felt needlessly complicated and like the privacy policy was being intentionally hidden behind several layers of user interaction.''} \textit{``It took forever to figure out where the privacy policy was.''}\\
S4 & Coping Strategies \& Desires & 25 (43\%) & \textit{``Confusing.''} \textit{``went to the website and found the privacy policy.''}\\
S4 & Design normalization \& Indifference & 11 (19\%) & \textit{``It operates just like any other Settings menu I've experienced.''} \textit{``Most privacy options are under settings/options and readily available.''}\\ \bottomrule
\multicolumn{4}{l}{\begin{tabular}[c]{@{}l@{}}\small \textit{Note. n=number of participants whose qualitative responses were coded under a given theme. Percentages (\%) are calculated based on the total number of} \\ \small \textit{participants in a Scenario (S): 62 (S1), 59 (S2), 61 (S3), 58 (S4), 64 (S5), 61 (S6), and 59 (S7).} \end{tabular}}
\end{tabular}%
}
\end{table}
\renewcommand{\arraystretch}{1}
\renewcommand{\arraystretch}{1.2}
\begin{table}[!t]
\caption*{Table 11 Continued. This codebook presents the codes and themes we synthesized from participants' responses to the open-ended question ``Drawing from your previous experience using [the VR application], what would be your thoughts and actions if you encountered this [VR design mechanism]?'' (Q9) and ``Please briefly explain the reasons...'' (Q10).}
\vspace{-3mm}
\label{tab:codebook-Q19thoughtsreactions-part2}
\centering
\resizebox{\textwidth}{!}{%
\begin{tabular}{@{}llcp{0.6\textwidth}@{}}
\toprule
\textbf{Scenario (S)} & \textbf{Themes} & \textbf{n (\%)} &\textbf{Example Quotes} \\ \midrule
S5 & Perceived Trust and Legitimacy & 38 (59\%) & \textit{``The application was allowed on Meta (owned by Facebook), I am sure the invasion of privacy can not be so bad that I would have to read it.''} \textit{``My thought were positive and I consent with this privacy form.''}\\
S5 & Concerns \& Reluctance & 27 (42\%) & \textit{``I did wonder about what they were using my personal information for, but I agreed.''} \textit{``It seemed to me that collecting eye movements and whatever else is kind of nerve-wracking.''}\\
S5 & Coping Strategies \& Desires & 16 (25\%) & \textit{``I felt it [the privacy policy] should've been on the start menu.''} \textit{``I figured if there's something truly bad in there I'd hear about it through some other avenue of the internet.''}\\
S5 & Design normalization \& Indifference & 36 (56\%) & \textit{``I'm pretty sure 99\% of people in VR would click on whatever they had to to make the text go away as fast as possible.''} \textit{``Most of these forms are required but are too long to sit down and read all of it.''}\\
S5 & Lack of Choice & 6 (9\%) & \textit{``Agreed to it in order to move forward.''} \textit{``There was a lot of legal words and definitions that confused me but if I didn't accept I wouldn't be able to proceed.''}\\
& & & \\
S6 & Perceived Trust and Legitimacy & 39 (64\%) & \textit{``I think it's a great way to personalize the game even further and really make it feel like you're the one in the VR simulation.''} \textit{``Positive because of the customizability and able to make it appear more real.''}\\
S6 & Concerns \& Reluctance & 1 (2\%) & \textit{``I customized my character and checked privacy settings to make sure nothing's shared without my consent.''}\\
S6 & Coping Strategies \& Desires & 10 (16\%) & \textit{``I would want more skin tones to pick from. I don't feel that the skin tones reflect mine (White).''} \textit{``I would like to have colors of skin such as blue or purple to make it more exciting and personalized.''}\\
S6 & Design normalization \& Indifference & 7 (11\%) & \textit{``It all seemed pretty normal to me.''} \textit{``This is an appropriate customization feature for the app.''}\\
S6 & Lack of Choice & 2 (3\%) & \textit{``Because I wanted to play the game and this was the first step.''}\\
& & & \\
S7 & Perceived Trust and Legitimacy & 10 (17\%) & \textit{``Makes sense if its for the system to work properly or smoothly.''} \textit{``They probably needed access to certain things for everything to work.''}\\
S7 & Concerns \& Reluctance & 31 (53\%) & \textit{``The program asked if I wanted to share certain information, I said no, and it seems like it's trying it's hardest to get me to change my mind without directly banning my gameplay for sharing the information that it wants me to share.''} \textit{``upset because it continued to happen and I didn't wanna give access to my info or data.''}\\
S7 & Coping Strategies \& Desires & 24 (41\%) & \textit{``I simply reinstalled the app.''} \textit{``I just restarted the whole thing...''}\\
S7 & Design normalization \& Indifference & 4 (7\%) & \textit{``It felt very invasive. It reminds me of the constant notifications I was getting to sign into my facebook to continue having access which I did.''} \textit{``I did not pay too much attention.''}\\
S7 & Lack of Choice & 12 (20\%) & \textit{``I was literally forced to allow requested information.''} \textit{``Sadly, I had to blindly accept these terms.''}\\ 
S8 & Perceived Trust and Legitimacy & 52 (91\%) & \textit{``It introduces option for me to manage my audio options, where I can decide to mute my audio as I please.''}\\
S8 & Coping Strategies \& Desires & 28 (49\%) & \textit{``A brighter layout would make me most likely looked at the board.''}\\
S8 & Design normalization \& Indifference & 4 (7\%) & \textit{``I would ignore and keep walking. Nobody is going to read a tutorial they aren't forced to read.''}\\
S8 & Concerns \& Reluctance & 2 (4\%) & \textit{``I'd be annoyed because I already know how to active my mic.''} \textit{``I'd be concerned about my privacy and would check to see if I'm open mic.''} \\
\bottomrule
\multicolumn{4}{l}{\begin{tabular}[c]{@{}l@{}}\small \textit{Note. n=number of participants whose qualitative responses were coded under a given theme. Percentages (\%) are calculated based on the total number of} \\ \small \textit{participants in a Scenario (S): 62 (S1), 59 (S2), 61 (S3), 58 (S4), 64 (S5), 61 (S6), 59 (S7), and 57 (S8).} \end{tabular}}
\end{tabular}%
}
\end{table}
\renewcommand{\arraystretch}{1}

\renewcommand{\arraystretch}{1.2}
\begin{table}[!t]
\caption{This codebook presents the codes and themes we synthesized from participants' responses to the open-ended question ``Please explain how [the selected] design element affected your decision during your own experience with this [VR design mechanism]?'' (Q15).}
\label{tab:codebook-Q33designelement}
\centering
\resizebox{\textwidth}{!}{%
\begin{tabular}{@{}llcp{0.6\textwidth}@{}}
\toprule
\textbf{Scenario} & \textbf{Theme} & \textbf{n (\%)} & \textbf{Example Quotes} \\ \midrule
S1 & Informed and Voluntary Decision & 37 (60\%) & \textit{``I like the freedom to decide how my personal data is used or what personal data is collected in the first place.''} \\
S1 & Persuasion and Interruption on decisions & 20 (32\%) & \textit{``It already has it checked. You would have to make a conscious effort to uncheck it. Most people wouldn't bother and would leave it marked.''} \\
S1 & Indifference, Irrelevance, Don't Know & 8 (13\%) & \textit{``UI elements were irrelevant in making my decision.''} \\
& & & \\
S2 & Informed and Voluntary Decision & 28 (47\%) & \textit{``Didn't want to enter it again.''} \textit{``The input field made the experience very simple and straight forward.''} \\
S2 & Persuasion and Interruption on decisions & 31 (53\%) & \textit{``The colors, textures and contrast of the background feeds my interest in using it.''} \textit{``I remember wanting to use a fake name.''}\\
S2 & Indifference, Irrelevance, Don't Know & 22 (37\%) & \textit{``Didn't affect me much.''} \textit{``Just was normal to me.''}\\
& & & \\
S3 & Informed and Voluntary Decision & 39 (64\%) & \textit{``It gave me the choices if I wanted to continue.''} \textit{``I'm glad it was presented clearly.''}\\
S3 & Persuasion and Interruption on decisions & 34 (56\%) & \textit{``when I put that headset on and got in sn I saw that realistic landscape and was already hooked.''} \textit{``they could be fake ads which those appeared to be because the photos alongside the ads looked too polished in my opinion.''} \\
S3 & Indifference, Irrelevance, Don't Know & 7 (11\%) & \textit{``I didn't notice it.''} \textit{``The appearance of how the information is displayed doesn't make a difference to me. My decision was based just on the information.''} \\
& & & \\
S4 & Informed and Voluntary Decision & 27 (47\%) & \textit{``The layout of the navigation button is quite easy to use.''} \textit{``It is easy to navigate.''} \\
S4 & Persuasion and Interruption on decisions & 13 (22\%) & \textit{``It wasnt clear.''} \textit{``It looks goofy and the privacy policy is tucked off to the side. Do not like the UI, never did.''} \\
S4 & Indifference, Irrelevance, Don't Know & 4 (7\%) & \textit{``Settings are a normal feature.''} \textit{``Every game has a main menu with a choice for settings.''}\\
& & & \\
S5 & Informed and Voluntary Decision & 17 (27\%) & \textit{``It made it easy for me to read and scan through the passage quickly.''} \textit{``It was clear to read and understandable.''}\\
S5 & Persuasion and Interruption on decisions & 4 (6\%) & \textit{``it doesnt it is very plain to see what you are agreeing to.''} \textit{``A wall of text placed in front me causes my eyes to look for the escape route so to speak.''}\\
S5 & Indifference, Irrelevance, Don't Know & 6 (9\%) & \textit{``I think it looked normal.''} \textit{``I scrolled and browsed it. I don't recall seeing anything unusual.''} \\
& & & \\
S6 & Informed and Voluntary Decision & 37 (61\%) & \textit{``It was easy for me to choose the gender I want.''} \textit{``they made it more realistic.''}\\
S6 & Persuasion and Interruption on decisions & 35 (57\%) & \textit{``I felt the urge that the hands should resemble mine.''} \textit{``I tried a couple options and went with what matched me closest based on the virtual representation of the hands.''}\\
S6 & Indifference, Irrelevance, Don't Know & 9 (15\%) & \textit{``It felt like a real hand, that's all.''} \textit{``I still would have chosen it.''}\\
\bottomrule
\multicolumn{4}{l}{\begin{tabular}[c]{@{}l@{}}\small \textit{Note. n=number of participants whose qualitative responses were coded under a given theme. Percentages (\%) are calculated based on the total number of} \\ \small \textit{participants in a Scenario (S): 62 (S1), 59 (S2), 61 (S3), 58 (S4), 64 (S5), 61 (S6), and 59 (S7).} \end{tabular}}
\end{tabular}%
}
\end{table}
\renewcommand{\arraystretch}{1}
\renewcommand{\arraystretch}{1.2}
\begin{table}[!t]
\caption*{Table 12 Continued. This codebook presents the codes and themes we synthesized from participants' responses to the open-ended question ``Please explain how [the selected] design element affected your decision during your own experience with this [VR design mechanism]?'' (Q15).}
\label{tab:codebook-Q33designelementpart2}
\centering
\resizebox{\textwidth}{!}{%
\begin{tabular}{@{}llcp{0.6\textwidth}@{}}
\toprule
\textbf{Scenario} & \textbf{Theme} & \textbf{n (\%)} & \textbf{Example Quotes} \\ \midrule
S7 & Informed and Voluntary Decision & 4 (7\%) & \textit{``I will try to address that as soon as possible to make it go away so I can continue uninterrupted.''} \\
S7 & Persuasion and Interruption on decisions & 22 (37\%) & \textit{``I think there should be a `dont show again' option. Users should not have to share privacy if they do not want to it does not affect the outcome of the apps performance in my opinion.''} \\
S7 & Indifference, Irrelevance, Don't Know & 1 (2\%) & \textit{``it reminds me that i can adjust it in the settings as a warning.''} \textit{``it's simple and seems like an easy fix.''}   \\ 
S8 & Informed and Voluntary Decision & 54 (95\%) & \textit{``Clearly explained the steps I needed to take to update my mic settings.''} \textit{``It told me where to go to find the option to mute my mic.''}\\
S8 & Persuasion and Interruption on decisions & - & - \\
S8 & Indifference, Irrelevance, Don't Know & 9 (16\%) & \textit{``I actually skip them. I have full knowledge of its use.''} \\
\bottomrule
\multicolumn{4}{l}{\begin{tabular}[c]{@{}l@{}}\small \textit{Note. n=number of participants whose qualitative responses were coded under a given theme. Percentages (\%) are calculated based on the total number of} \\ \small \textit{participants in a Scenario (S): 62 (S1), 59 (S2), 61 (S3), 58 (S4), 64 (S5), 61 (S6), 59 (S7), and 57 (S8).} \end{tabular}}
\end{tabular}%
}
\end{table}
\renewcommand{\arraystretch}{1}


\end{document}